%% file: IEEE_Template_v1.tex
\documentclass[journal,draftclsnofoot,onecolumn,12pt]{IEEEtran}

\usepackage{amsthm,amssymb,graphicx,multirow,amsmath,color,amsfonts}
\usepackage[update,prepend]{epstopdf}
\usepackage[noadjust]{cite}
\usepackage[utf8]{inputenc} 
\usepackage{tikz}
\usetikzlibrary{arrows,calc}		
\usepackage{bbm} 
\usepackage{pdfpages}
\usepackage{tabulary}
\usepackage{multirow}
\usepackage{comment}
\usepackage[ruled,vlined]{algorithm2e}
\usepackage{etoolbox}
\usepackage{hyperref}

\usepackage[font=small,labelfont=bf]{caption}
\usepackage{subcaption}
\makeatletter
\renewcommand{\SetKwInOut}[2]{%
  \sbox\algocf@inoutbox{\KwSty{#2}\algocf@typo:}%
  \expandafter\ifx\csname InOutSizeDefined\endcsname\relax
    \newcommand\InOutSizeDefined{}\setlength{\inoutsize}{\wd\algocf@inoutbox}%
    \sbox\algocf@inoutbox{\parbox[t]{\inoutsize}{\KwSty{#2}\algocf@typo:\hfill}~}\setlength{\inoutindent}{\wd\algocf@inoutbox}%
  \else
    \ifdim\wd\algocf@inoutbox>\inoutsize%
    \setlength{\inoutsize}{\wd\algocf@inoutbox}%
    \sbox\algocf@inoutbox{\parbox[t]{\inoutsize}{\KwSty{#2}\algocf@typo:\hfill}~}\setlength{\inoutindent}{\wd\algocf@inoutbox}%
    \fi%
  \fi
  \algocf@newcommand{#1}[1]{%
    \ifthenelse{\boolean{algocf@inoutnumbered}}{\relax}{\everypar={\relax}}%
    {\let\\\algocf@newinout\hangindent=\inoutindent\hangafter=1\parbox[t]{\inoutsize}{\KwSty{#2}\algocf@typo:\hfill}~##1\par}%
    \algocf@linesnumbered
  }}%
\makeatother

\SetKwInOut{KwInput}{Input}
\SetKwInOut{KwOutput}{Output}

\include{notation}

\allowdisplaybreaks 
\IEEEoverridecommandlockouts
\begin{document}
\graphicspath{{./Figures/}}
\title{
Robust Optimization of RIS in Terahertz under Extreme Molecular Re-radiation Manifestations
}
\author{\thanks{}}
\author{ Anish Pradhan,
Mohamed A. Abd-Elmagid, Harpreet S. Dhillon and Andreas F. Molisch
\thanks{A. Pradhan, M. A. Abd-Elmagid, and H. S. Dhillon  are with Wireless@VT, Department of ECE, Virginia Tech, Blacksburg, VA, USA (email: \{pradhananish1, maelaziz, hdhillon\}@vt.edu). A. F. Molisch is with the Wireless Devices and Systems Group, Ming Hsieh Department of Electrical and Computer Engineering, University of Southern California, Los Angeles, CA, USA (email: molisch@usc.edu).  This work was supported by U.S. National Science Foundation under Grant ECCS-2030215.

This paper was presented in part at the IEEE Globecom 2021, Madrid, Spain \cite{gcs}.
 } 
 \vspace{-5mm}
}

\maketitle

\begin{abstract}

Terahertz (THz) communication signals are susceptible to severe degradation because of the molecular interaction with the atmosphere in the form of subsequent absorption and re-radiation. Recently, reconfigurable intelligent surface (RIS) has emerged as a potential technology to assist in THz communications by boosting signal power or providing virtual line-of-sight (LOS) paths. However, the re-radiated energy has either been modeled as a non-line-of-sight (NLOS) scattering component or as additive Gaussian noise in the literature. Since the precise characterization is still a work in progress, this paper presents the first comparative investigation of the performance of an RIS-aided THz system under these two extreme re-radiation models. In particular, we first develop a novel parametric channel model that encompasses both models of the re-radiation through a simple parameter change, and then utilize that to design a robust block-coordinate descent (BCD) algorithmic framework which maximizes a lower bound on channel capacity while accounting for imperfect channel state information (CSI). In this framework, the original problem is split into two sub-problems: a) receive beamformer optimization, and b) RIS phase-shift optimization. As the latter sub-problem (unlike the former) has no analytical solution, we propose three approaches for it: a) semi-definite relaxation (SDR) (high complexity), b) signal alignment (SA) (low complexity), and c) gradient descent (GD) (low complexity). The time complexities associated with the proposed approaches are explicitly derived. We analytically demonstrate the limited interference suppression capability of a passive RIS by deriving the stationary points of signal-to-interference and noise ratio (SINR) of a one-element RIS system with one interferer. Our numerical results also demonstrate that slightly better throughput is achieved when the re-radiation manifests as scattering. 

\end{abstract}


\section{Introduction} \label{sec:intro}
With the standardization of 5G new radio (NR), it is now well-accepted that the traditional sub-6 GHz spectrum by itself is not sufficient to meet the ever-expanding network demands in the near future \cite{tripathi2021millimeter,enough5G}. This has led to the pursuit of utilizing higher frequency bands, which ultimately resulted in the recent commercialization of mmWave communication. However, with the advent of extended reality (xR) technologies, even higher data rates - up to 1 Tbit/s - for which mmWave bandwidths are not sufficient anymore, are required \cite{tataria20216g}. The reason is that the xR ecosystem imposes very stringent requirements on throughput of the wireless communication technologies sustaining it. Once realized, the xR applications are expected to revolutionize many industry sectors, including, but not limited to, healthcare, entertainment, and eCommerce. To support such applications, there has been a recent interest in exploring the possibility of utilizing the THz (0.1-10 THz) spectrum, which lies above the mmWave band \cite{tripathi2021millimeter}. Recent breakthroughs in the research of high-power THz sources \cite{graphenesource, GaNsource} have further increased the viability of utilizing this spectrum. 

However, THz communication links are highly susceptible to blockages, both by static objects, and by dynamic objects including the users operating the VR \cite{VRTHz3}. Static blockages consistently prevent suitable quality of experience (QoE), while dynamic blockages result in a sudden decrease in throughput and are detrimental to the immersion of xR. A further impairment for THz signals is the molecular re-radiation that can manifest as either noise or NLOS component of the signal. 

Inspired by the recent standardization efforts by various organizations, a potential solution is to deploy the emerging RIS technology that can create virtual LOS links to enhance throughput in situations where direct LOS links are blocked. Yet, the integration of RIS with THz communication links presents the following challenges: a) accurate characterization of the molecular re-radiation, b) channel estimation issues due to RIS, and c) subsequent operation of RIS under imperfect CSI. To overcome these challenges, we first develop a parametric THz channel model in this paper that captures both manifestations of re-radiation, and then use that model to present a novel alternating RIS optimization framework, where the RIS phase shift and receive beamformer are jointly optimized.

\subsection{Background and Prior Art}
We will now discuss in more detail each of the aforementioned challenges associated with integrating RIS with THz communication links. We start this discussion with the challenge of an accurate characterization of the molecular re-radiation, which is the less understood and is also the prime motivation behind this work. In most of our typical communication scenarios, water vapor is one of the primary constituents in the molecular makeup of the wireless medium. Since water molecule, like many other atmospheric molecules \cite{discussion}, has many rotational absorption lines through the THz band \cite{Yang:14}, these molecules are highly susceptible to being excited by the THz communication signals. In particular, the transmitted EM wave causes molecular absorption by exciting the molecules from lower to higher energy states. These higher energy molecules re-radiate absorbed energy in a similar frequency range while returning to the ground state. For many decades, the process of such atomic and molecule re-radiation has been referred to as {\em radiation trapping} in the physics literature \cite{molisch1998radiation}. In the existing THz literature, this re-radiation often manifests as additive Gaussian noise based on \textit{sky-noise} models \cite{discussion, jornet_2012}. This is an approximation that results from the fundamental difference of the physical phenomena dictating the two \cite{discussion}. To our knowledge, no measurement studies have adequately supported this model until now. Furthermore, there is some support in the literature \cite{discussion, Harde1, Harde2} for describing this phenomenon as scattering, with the presence of multiple scattered copies of the signal due to re-radiation. Note that this scattering could potentially result in delay dispersion \cite{discussion} and frequency dispersion \cite{molisch1998radiation}, which is beyond the scope of this paper. Following the scattering assumption, \cite{molabs} recently characterized the THz channel as a Rician channel, with the Rician factor computed from the molecule absorption coefficient. In the literature, both manifestations (i.e., as noise and as scattering) have been employed separately, and determining the prevalence of each effect is difficult. To be more specific, the exact effect will most likely exist as a combination of these two extreme situations. There is no way to accurately define the exact effect without comprehensive measurement studies. Given this, one of the reasonable things to do with the current information is to explore the two extreme circumstances and quantify their influence on the RIS performance which is what we do in this paper.

Due to the peculiarities of the THz links, the techniques developed in the beamforming literature on RIS \cite{wu2019beamforming,RISCR,ye2020joint,activeRIS,sp,ee} in the sub-6 GHz spectrum (that deal with the joint optimization of RIS phase-shifts and receiver beamformer) cannot be trivially extended. The interplay between RIS and THz band has been recently studied in \cite{VRTHz3,VRTHz2,VRTHz1,risthzcm,news,twc,leo}. The authors of \cite{VRTHz2} proposed a sub-optimal search method to optimize RIS discrete phase-shifts while the authors of \cite{VRTHz1} jointly optimized the RIS location, phase-shift and THz sub-bands to improve system performance. A deep reinforcement learning-based algorithm has been used to optimize the reliability and rate for RIS-operated virtual reality systems in the THz band \cite{VRTHz3}. A physically consistent near-field channel model for RIS-THz systems was developed in \cite{risthzcm} while the secrecy rate for an RIS-aided THz system was optimized in \cite{news}. A particle swarm optimization-based method with limited channel estimation was used to optimize the RIS phase shifts in a THz band \cite{twc}. The error performance of an RIS-assisted low earth orbit satellite network has been analyzed in \cite{leo}. However, the above prior works studying RIS in the THz band neglected either the two possible manifestations of the molecular re-radiation, or the cumulative effect of this re-radiation along with the RIS configuration on the receiver noise \cite{VRTHz2,VRTHz1,VRTHz3,risthzcm,news,twc,leo}. Further, they did not account for the natural challenge of imperfect CSI resulting from the passive nature of RIS elements, and non-cooperation from the interfering nodes.

The robust optimization of RIS-aided THz systems (against the imperfect CSI) has only been considered in a handful of recent works \cite{rob1,rob2,rob3}. These works used semidefinite programming (SDP) techniques while ignoring the peculiar characteristics of RIS-THz integration. Such techniques suffer from high computational cost that decreases the energy efficiency of the network, and hence defeat the purpose of low-cost RISs \cite{lowc}. We bridge this gap by developing a parametric THz channel model that accounts for both assumptions of re-radiation, as well as three BCD-based joint optimization approaches of varying complexity for the proposed channel model under imperfect CSI. We use a lower bound on the channel capacity as our objective as the exact channel capacity is unknown for considering interference in our system model \cite{massivemimobook}. Different from \cite{rob1,rob2}, the achievable throughput expression in our objective function assumes that the receiver only has access to imperfect CSI, which reflects the reality more precisely. Our objective function is also consistent with the discussion of the uplink spectral efficiency under imperfect CSI in \cite[eq. (4.1)]{massivemimobook}. To the best of our knowledge, no comparative study exists that analyzes a jointly-optimized multi-antenna system in a THz environment with two extreme assumptions regarding molecule re-radiation. 

\subsection{Contributions}
We study an RIS-aided THz system setting that consists of a single-stream transmitter (Tx) communicating with an RIS-aided multi-antenna receiver (Rx) in the THz band in the presence of potentially multiple single-stream interferers. For this setup, our objective is to jointly optimize the RIS's phase shift and receive beamformer while assuming imperfect CSI knowledge. Our key contributions in this paper are listed next.

\textit{A novel parametric THz channel.} We propose a new parametric THz channel model that accounts for the following two extreme manifestations of re-radiation in the THz spectrum through a single parameter change: a) re-radiation is assumed as Gaussian noise, and b) re-radiation is assumed as an NLOS component of the signal. We also characterize the cumulative effect of molecular re-radiation and the RIS configuration utilizing this parameter.

\textit{Three robust BCD algorithms.} We formulate an optimization problem in which we jointly optimize the RIS's phase shift vector and receive beamformer vector with the objective of maximizing a lower bound on the channel capacity. Due to the coupling between the two sets of optimization variables (i.e., the RIS's phase shift vector and receive beamformer vector) in its objective function, the formulated problem turns out to be non-convex, and hence its global optimal solution cannot be obtained using standard convex optimization techniques. Because of that, we aim to obtain an efficient solution through the BCD algorithm. In this algorithm, we split the original problem of two sets of optimization variables into the following two sub-problems of one set of variables each: a) receive beamforming vector optimization problem, and b) RIS's phase shift optimization problem. These sub-problems are then solved in an alternative manner until they converge to an efficient solution of the original problem. As the latter sub-problem does not have a closed-form solution unlike the former, we propose three algorithms of varying complexity to solve the RIS sub-problem. First, we propose a conventional SDR approach as a baseline. Due to the high time complexity of the SDR approach, we then present the SA approach for its speed, where the expected receive signal strength is maximized rather than the original objective function. This approach provides a good sub-optimal solution when the interference power in the network is low. However, in a network with a moderate amount of interference, we can achieve better performance without sacrificing any speed by utilizing the gradient descent algorithm, which is our third proposed approach. These approaches consider the direct links of both users and interferers under imperfect CSI. Our objective function also caters to the non-robust counterpart by simply assuming no error. Finally, the time complexities associated with these approaches are explicitly derived.

\textit{System design insights.} We analytically characterize the performance loss associated with the SA solution by deriving the stationary points of a one-element RIS-aided system with one interferer. This allows us to demonstrate that the passive RIS has limited capability to suppress interference when the direct link of the interferer is much stronger than the reflected link. Multiple system design insights can also be drawn from the numerical results. For example, our numerical results reveal that when re-radiation manifests as scattering, the corresponding throughput of the optimized system is slightly higher than when it manifests as noise. They also show that the gap in performance of the two cases depends on the visibility of the interferer direct links and frequency. Under perfect CSI, throughput is shown to increase linearly and logarithmically with the increasing number of RIS elements and Rx antennas, respectively. Further, we do not observe much penalty in performance by assuming the nature of molecular re-radiation in the optimization method incorrectly whenever perfect CSI is available. The results also show that the proposed robust algorithms perform better than the non-robust counterparts under imperfect CSI. In particular, our results demonstrate that the BCD-GD algorithm is superior in terms of the runtime and SER performance.

\subsubsection*{Notations}
The scalar, vector and matrix are denoted by $x$, $\bf x$ and $\bf X$, respectively. All the vectors are column vectors unless stated explicitly. For a matrix $\bf X$, ${\bf X}^T$, ${\bf X}^H$, ${\rm Tr}\left({\bf X}\right)$, $[{\bf X}]_{i,j}$, ${\rm Re}\left({\bf X}\right)$  and ${\bf X}\succeq0$ denote its transpose, conjugate transpose, trace, $(i,j)$-th element, real part and positive semidefiniteness, respectively. The operation ${\rm vec}({\bf X})$ results in a vector with every element of ${\bf X}$. For a vector $\bf x$, ${\rm diag}\left({\bf x}\right)$ denotes the diagonal matrix with the elements of $\bf x$ as its diagonal elements. The element-wise product is denoted by $\odot$. The distribution of a standard complex normal random variable is denoted by $\mathcal{CN}(0,1)$.

\section{System Model} \label{sec:SysMod}

\begin{figure}
    \centering
    \includegraphics[width=0.5\linewidth]{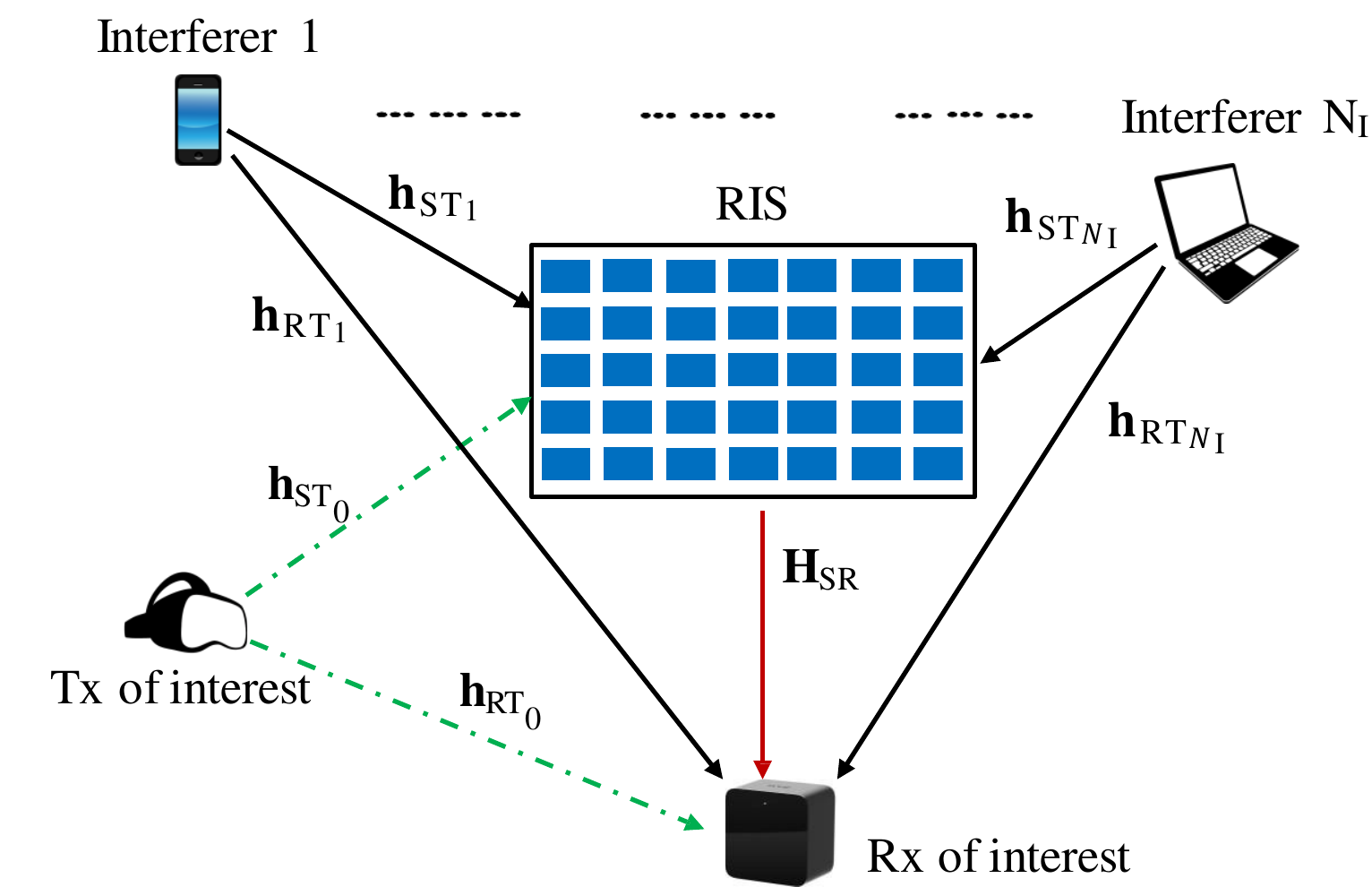}
    \caption{System model for the RIS-assisted communication in THz.}
    \label{fig:sysmodel2}
\end{figure}

We consider an RIS-aided THz system setup inspired by an indoor xR scenario with multiple Tx-Rx pairs communicating simultaneously in the same THz frequency band. The Tx of interest (${\rm Tx}_0$) is assumed to be a single-stream device without any active beamforming capabilities (like a VR/AR user equipment with a small form factor) \cite{vrtx}, while the Rx of interest (${\rm Rx}_0$) is considered to have multiple antennas (similar to a mobile edge computing server) \cite{VRTHz3} and is assisted by a passive RIS. The multi-antenna ${\rm Rx}_0$ has $N_R$ receive antennas, while the RIS has $N$ elements. Additionally, there are $N_I$ co-channel single-stream users that interfere at the direction of ${\rm Rx}_0$. It is assumed that each of the communicating devices can have two paths to ${\rm Rx}_0$, one link coming directly from the Tx, and another link reflected from the RIS.
The system model is illustrated in Fig. \ref{fig:sysmodel2}. Note that the $N_I$ interfering users are communicating with their own Rxs. These Rxs do not affect our analysis and are therefore not included in Fig. \ref{fig:sysmodel2}. Both RIS and ${\rm Rx}_0$ are assumed to be uniform rectangular arrays (URAs) with half-wavelength spacing to sufficiently decrease mutual coupling \cite{mutualc}. The array factor of a general URA with $N_0$ elements is defined as follows \cite{massivemimobook}:
\begin{align}
 \mathbf{a}_{N_0}(\varphi,\theta,{\bf U})&=\begin{bmatrix}
e^{j{\bf k}(\varphi,\theta)^T{\bf u}_1}, \hdots,e^{j{\bf k}(\varphi,\theta)^T{\bf u}_{N_0}}
\end{bmatrix}, \label{eq:afactor}
\end{align}
where ${\bf k}(\varphi,\theta)=\frac{2 \pi}{\lambda}\left[\cos (\theta) \cos (\varphi) ,
\cos (\theta) \sin (\varphi) ,
\sin (\theta)\right]^T$ is the wave vector, $\lambda$ denotes the wavelength, ${\bf u}_i$ denotes the vector of Cartesian co-ordinates of the $i$-th URA element, and ${\bf U}$ denotes the matrix $[{\bf u}_1,{\bf u}_2,\hdots,{\bf u}_{N_0} ]$. The azimuth angle $\varphi\in[-\pi,\pi)$ is measured from the positive x-axis and the elevation angle $\theta\in[-\frac{\pi}{2},\frac{\pi}{2})$ is measured from the x-y plane. The array is assumed to be on the positive y-z plane with the origin as the reference. Note that the array factor is defined as a row vector.

\subsection{Terahertz Channel Model}

Owing to the molecular absorption phenomena in the THz band, a fraction $1-\tau(f,d)$ of the propagating signal is absorbed and re-radiated by the media. The remaining fraction $\tau(f,d)=e^{-k(f)d}$ is termed as the transmittance of the channel, where $k(f),f$ and $d$ denote the molecular absorption coefficient, the operating frequency, and the link distance, respectively. Further, we use a simple LOS channel model that is valid for 200-450 GHz \cite{SimpleTHz} for the molecular absorption coefficient calculation. The equation dictating the value of $k(f)$ is:
\begin{align}
  k(f)=\sum_i y_i(f,\mu)+g(f,\mu),
\end{align}
where $\mu$ denotes the volume mixing ratio of water vapor, $y_i(f,\mu)$ denotes the absorption coefficient for the $i$-th absorption line, and the polynomial function $g(f,\mu)$ is an equalization factor. The volume mixing ratio $\mu$ can be calculated through the following equation:
\begin{align}
    \mu=\frac{\phi}{100} \frac{p_{w}(T, p)}{p},
\end{align}
where $\phi$ and $p$ denote the relative humidity and pressure, respectively. The function $p_{w}(T, p)$ can be described by the Buck equation \cite{buck}:
\begin{align}
    p_{w}(T, p)=6.1121\left(1.0007+3.46 \times 10^{-6} p\right) \exp \left(\frac{17.502 T}{240.97+T}\right),
\end{align} where the pressure $p$ is in hectopascals and the temperature $T$ is in Celsius. Next, we express $\{y_i(f, \mu)\}$ and $g(f,\mu)$ as described in \cite{SimpleTHz}:
\begin{align*}
&y_{1}(f, \mu) =\frac{A(\mu)}{B(\mu)+\left(\frac{f}{100 c}-p_{1}\right)},\quad y_{2}(f, \mu) =\frac{C(\mu)}{D(\mu)+\left(\frac{f}{100 c}-p_{2}\right)},
\\& y_{3}(f, \mu) =\frac{E(\mu)}{F(\mu)+\left(\frac{f}{100 c}-p_{3}\right)},\quad  g(f, \mu)=\frac{\mu}{0.0157}\left(q_{1} f^{4}+q_{2} f^{3}+q_{3} f^{2}+q_{4} f+q_{5}\right),
\end{align*}

where the frequency $f$ is in Hertz and the various inner entities are described below:
\begin{align*}
A(\mu)=0.2251 \mu(0.1314 \mu+0.0297),\quad  
B(\mu)=(0.4127 \mu+0.0932)^{2}, \\
C(\mu)=2.053 \mu(0.1717 \mu+0.0306),\quad  
D(\mu)=(0.5394 \mu+0.0961)^{2}, \\
E(\mu)=0.177 \mu(0.0832 \mu+0.0213),\quad  
F(\mu)=(0.2615 \mu+0.0668)^{2}, \\
G(\mu)=2.146 \mu(0.1206 \mu+0.0277),\quad  
H(\mu)=(0.3789 \mu+0.0871)^{2}.
\end{align*}

The constants used are: $p_{1}=10.84 \mathrm{~cm}^{-1}, p_{2}=12.68 \mathrm{~cm}^{-1}, p_{3}=14.65 \mathrm{~cm}^{-1}, p_{4}=14.94 \mathrm{~cm}^{-1}$, $q_{1}=8.495 \times 10^{-48}, q_{2}= -9.932 \times 10^{-36}, q_{3}=4.336 \times 10^{-24}, q_{4}=-8.33 \times 10^{-13},
\text { and } q_{5}=5.953 \times 10^{-2}$. The absorption lines $y_{1}(f, \mu), y_{2}(f, \mu), y_{3}(f, \mu)$, and $y_{4}(f, \mu)$ represent the absorption peaks at center frequencies $325$ GHz, $380$ {GHz}, $439$ {GHz}, and $448$ {GHz}, respectively.

Recall from Section~\ref{sec:intro} that the absorbed power is emitted again in the same band, and there are two extreme modeling assumptions regarding how this molecular re-radiation manifests. Unlike the forward scattering we witness in visible light owing to tiny particles, this re-radiation occurs virtually isotropically, and the energy is spread in all directions for the THz band \cite{vanExter:89}. However, it is commonly assumed that the whole absorbed power is available at the Rx node through re-radiation \cite{jornet_2012,molabs}. This result follows from the assumptions that a single absorption/re-radiation event occurs throughout the entire propagation route, and that all re-radiated power is directed in the direction of the Rx node, despite the fact that this is never explicitly stated. In our paper, we also apply these assumptions. The assumptions of the two extremes of molecular re-radiation are provided next for a more systematic exposition.
\begin{assumption} \label{assumption:noise}
Molecular re-radiation is modeled as additive Gaussian noise. 
\end{assumption}

\begin{assumption} \label{assumption:scatter}
Molecular re-radiation is modeled as a scattering event where the affected channel response includes an NLOS component. 
\end{assumption}
The re-radiated signal, which has $1-\tau(f,d)$ of the total signal power appears as either additive noise or as the NLOS component. Because of the nature of their modeling, the two assumptions are at opposite extremes of the spectrum, with reality falling somewhere in the middle. The Rician factor $K_d$ in Assumption \ref{assumption:scatter} for distance $d$ is
\begin{align}
    K_d=\frac{\text{Power of the LOS channel}}{\text{Power of the NLOS channel}}=\frac{\tau(f,d)}{1-\tau(f,d)}.
\end{align}

The two assumptions can also be interpreted as resulting from different possible receiver structures. Firstly, we have to recognize that radiation trapping generally leads to both delay and frequency dispersion, with the former arising from the finite lifetime of the excited molecular states, and the latter arising from (partial or complete) frequency redistribution occurring about relaxation of the excited state into the ground state \cite{molisch1998radiation}. If the frequency redistribution is over a band that is larger than the transmission bandwidth, part of the absorbed energy will be completely lost, while the remainder stays in the considered band. Depending on the amount of dispersion and its time variance, as well as the considered modulation and coding scheme, the scattered energy can be exploited for signal detection, or act as interference. For this reason, this paper considers the two limiting cases of ''fully useful" and ''fully noise". 

By including a variable $\zeta$ in the standard Rician channel model \cite{molabs}, we unify both Assumptions \ref{assumption:noise} and \ref{assumption:scatter} concerning the re-radiation process for analytical convenience. In the first scenario, the unified model should only include the LOS component since the re-radiation signal appears as Gaussian noise, whereas the channel model in the second case should include both the LOS and NLOS components because re-radiation appears as scattering. In other words, the models corresponding to Assumptions \ref{assumption:noise} and \ref{assumption:scatter} can be recovered from the unified model by setting $\zeta = 1$ and $\zeta = 0$, respectively. Further details on the modified Rician channel with the $\zeta$ parameter are provided next.

The channel response between $\mathrm{X}$ and $\mathrm{Y}$ is denoted by $\mathbf{h}_{\mathrm{XY}}$ or $\mathbf{H}_{\mathrm{XY}}$ (depending on the number of antenna elements at $\mathrm{X}$ and $\mathrm{Y}$), and can be expressed as:
\begin{align}
    &\mathbf{h}_{\mathrm{XY}}\!=\!\left(\sqrt{\frac{K_{d}}{K_{d}+1}}\mathbf{f}_{\mathrm{LOS}}e^{j\varpi}+\sqrt{\frac{1-\zeta}{K_{d}+1}}\Tilde{\mathbf{h}}_{{\rm XY}}\right)\frac{c}{4\pi f d}, \label{eq:directchannel} \\
    &\mathbf{H}_{\mathrm{XY}}\!=\!\left(\sqrt{\frac{K_{d}}{K_{d}+1}}\mathbf{F}_{\mathrm{LOS}}e^{j\varpi}+\sqrt{\frac{1-\zeta}{K_{d}+1}}\Tilde{\mathbf{H}}_{{\rm XY}}\right)\frac{c}{4\pi f d}, \label{eq:commonchannel}
\end{align}
where $\varpi$ denotes a random phase uniformly distributed in $[-\pi,\pi)$, $\mathrm{X=S\text{ or }R}$ represents RIS or the ${\rm Rx}_0$, and $\mathrm{Y=R}\text{ or }\mathrm{T}_i$ represents the the ${\rm Rx}_0$ or the $i$-th Tx. The distance between $\mathrm{X}$ and $\mathrm{Y}$ is denoted by $d$ while $f$ is the transmission frequency. The geometric channel related to the LOS path is denoted by $\mathbf{f}_\mathrm{LOS}$ (or $\mathbf{F}_\mathrm{LOS}$) and the NLOS counterpart is denoted by $\Tilde{\mathbf{h}}_{{\rm XY}}$ (or $\Tilde{\mathbf{H}}_{{\rm XY}}$) which is a complex vector (or matrix) with each entry being an independent and identically distributed (i.i.d.) complex normal random variable with zero mean and unit variance. Note that, when $\zeta=1$, the NLOS component vanishes and $\frac{K_{d}}{K_{d}+1}=\tau(f,d)$ corresponds to the transmittance, as expected. The distances and fixed channels associated with the different combinations of $\mathrm{X}$ and $\mathrm{Y}$ in the system model are compiled in Table \ref{tab:xy}. We define the following stacked channels for notational ease:
\begin{align}
&\mathbf{Z}_{i}=\mathbf{H}_\mathrm{SR}\mathrm{diag}\left(\mathbf{h}_{\mathrm{ST}_{i}}\right),
&\mathbf{H}_{i}=[\mathbf{Z}_{i}~ \mathcal{I}_i\mathbf{h}_{\mathrm{RT}_{i}}],~ \forall i\in [0,{N}_{I}], \label{eq:stacked}
\end{align}
where $\mathcal{I}_i$ is an indicator function that either takes $0$ or $1$ depending on the visibility of the direct link with probability $P_{L_i}$.
\renewcommand{\arraystretch}{1.3}
\begin{table}[]
\begin{center}
\begin{tabular}{ |c||c|c|c| } 
 \hline
 $\mathrm{XY}$ & $\mathrm{RT}_i$ & $\mathrm{ST}_i$ & ${\rm SR}$\\ 
 \hline
 $d$ & $d_i$ & $d_{\gamma_i}$ & $d_\alpha$ \\ 
 \hline
 $\mathbf{f}_\mathrm{LOS} \text{ or } \mathbf{F}_\mathrm{LOS}$ & $\mathbf{a}_{N_R}^H({\varphi}_{R,i},{\theta}_{R,i},{\bf U}_{BS})$ & $\mathbf{a}_{N}^H({\varphi}_{S,i},{\theta}_{S,i},{\bf U}_{RIS})$ & $\mathbf{a}_{N_R}^H({{\varphi}_{\alpha},\theta}_\alpha,{\bf U}_{BS})\mathbf{a}_{N}({\varphi}_{\beta},{\theta}_\beta,{\bf U}_{RIS})$ \\ 
 \hline
\end{tabular}
\end{center} 
\caption{\label{tab:xy}Channel Notations.}
\end{table}
\subsection{Imperfect CSI Model}
In a practical wireless system, wireless channels need to be estimated before reliable communication links can be established. Unlike the estimation of channels between communication nodes, the estimation of non-cooperative interferer channels is particularly challenging and error-prone. We use an additive error model \cite{imcsi1,imcsi2} to model this imperfect reflected channel as follows:
\begin{align}
    {\bf Z}_i=\hat{\bf Z}_i+{\bf \Delta}_i,
\end{align}
where the true channel is ${\bf Z}_i$, the estimated channel is $\hat{\bf Z}_i$, and the elements of the error matrix ${\bf \Delta}_i$ are i.i.d. as $\mathcal{CN}({0,\rho_i^2})$. Similarly, the error vector for the imperfect direct channel is $\boldsymbol{\delta}_i$ whose every element is i.i.d as $\mathcal{CN}({0,\rho_i'^2})$. The imperfect direct channel is expressed as follows:
\begin{align}
    {\bf h}_{{\rm RT}_i}=\hat{\bf h}_{{\rm RT}_i}+\boldsymbol{\delta}_i.
\end{align}

Now, estimated stacked channels are defined below:
\begin{align}
    &\hat{\mathbf{H}}_{i}=[\hat{\mathbf{Z}}_{i}~ \mathcal{I}_i\hat{\mathbf{h}}_{\mathrm{RT}_{i}}],~ \forall i\in [0,{N}_{I}]. \label{eq:stackedestimated}
\end{align}

We consider $\{\rho_i,\rho'_i\}$ to be parameters in our model that control the extent of the uncertainty in our channel estimates. Note that the channel is perfect when these parameters are set to zero. Larger values of these parameters denote worse CSI quality.

\subsection{Signal Model}
If the signal $x_{i}$ of power $\mathrm{E}[|x_i|^2]=P_i$ is transmitted by the $i$-th Tx, the received signal at ${\rm Rx}_0$ is expressed as \eqref{eq:signalmodel0}:
\begin{align}
    \mathbf{y}=&(\mathbf{h}_{\mathrm{RT}_0}+\mathbf{H}_\mathrm{SR}\mathrm{diag}\left(\mathbf{h}_{\mathrm{ST}_0}\right)\mathbf{\boldsymbol{\theta}})x_0+
    \sum\limits_{i=1}^{N_I}(\mathbf{h}_{\mathrm{RT}_{i}}+\mathbf{H}_\mathrm{SR}\mathrm{diag}\left(\mathbf{h}_{\mathrm{ST}_{i}}\right)\mathbf{\boldsymbol{\theta}}){x}_i + \mathbf{w}, \label{eq:signalmodel0}
\end{align}
where $ \mathbf{\boldsymbol{\theta}} = [e^{j\varphi_1}\,\hdots \,e^{j\varphi_N}]^T$ is the RIS configuration vector, $\varphi_n \in [0,2\pi]$ is the $n$-th entry of the vector $\boldsymbol{\varphi}$ for all $n\in\{1,\hdots,N\}$ and denotes the $n$-th element's reflection coefficient, and $\mathbf{w}$ denotes the additive Gaussian noise with variance $\sigma_w^2 + \zeta\sigma_m^2$. The variance terms $\sigma_w^2$ and $\zeta\sigma_m^2$ represent the thermal noise and molecular re-radiation noise, respectively. Molecular re-radiation noise under Assumption \ref{assumption:noise} can be calculated as: $\sigma_m^2=\sum\limits_{i=0}^{N_I}\sigma_{m,i}^2$, where $\sigma_{m,i}^2$ is the molecular re-radiation noise due to the $i$-th ${\rm Tx}$.
Note that, as this $\zeta=0$ conforms to Assumption \ref{assumption:scatter}, the molecular re-radiation noise variance disappears and manifests as fading.

Now, we have all the information to make the signal model more compact with the stacked channel structure (\ref{eq:stacked}). Using that, the received signal is rewritten as:
\begin{align}
\mathbf{y}&=\mathbf{H}_0\mathbf{\boldsymbol{\theta}}_0x_0+\sum\limits_{i=1}^{N_I}\mathbf{H}_{i}\mathbf{\boldsymbol{\theta}}_0x_i + \mathbf{w},  \label{eq:signal}
\end{align}
where ${\bf \boldsymbol{\theta}}_0=[{\bf \boldsymbol{\theta}}^T ~ 1]^T$.
Now, we multiply the received signal with the receive beamformer $\mathbf{u}^H$ from the left and express the resulting received signal as:
\begin{align}\label{eq:26}
\mathbf{u}^H\mathbf{y}&=\mathbf{u}^H\hat{\mathbf{H}}_{0}\mathbf{\boldsymbol{\theta}}_0x_0+\mathbf{u}^H\sum\limits_{i=1}^{N_I}\hat{\mathbf{H}}_{i}\mathbf{\boldsymbol{\theta}}_0x_i+\mathbf{u}^H\sum\limits_{i=0}^{N_I}\left(\mathbf{\Delta}_i\mathbf{\boldsymbol{\theta}}+\mathcal{I}_i\boldsymbol{\delta}_i\right)x_i + \mathbf{u}^H\mathbf{w}.
\end{align}

While the first term in (\ref{eq:26}) represents our desired signal, the last three terms denote interference, channel estimation errors, and noise, respectively. As the exact channel capacity is unknown for the interference channel, we provide a well-known lower bound on the channel capacity $C_{\rm sys}$ based on the discrete memoryless interference channel \cite[Corollary 1.3]{massivemimobook}:
\begin{align}
   C_{\rm sys} \geq \log_2\left(1+\gamma(\mathbf{u},\boldsymbol{\varphi})\right), \label{eq:ergocap}
\end{align}
such that the SINR term $\gamma(\mathbf{u},\boldsymbol{\varphi})$ can be expressed as follows:
\begin{align}
&\gamma(\mathbf{u},\boldsymbol{\varphi})=\frac{P_0|\mathbf{u}^H\hat{\mathbf{H}}_0\mathbf{\boldsymbol{\theta}}_0|^2}{\sum\limits_{i=1}^{N_I}P_i|\mathbf{u}^H \hat{\mathbf{H}}_{i}\mathbf{\boldsymbol{\theta}}_0|^2+\sum\limits_{i=0}^{N_I}P_i\mathbf{u}^H{\bf C}_{e_i}\mathbf{u}+\sigma_w^2+\zeta\sigma_{m}^2},\label{eq:SINRexpression}
\end{align}
where $\{{\bf C}_{e_i}\}$ are the co-variance matrices for the estimation errors and can be calculated as follows:
\begin{align}
    {\bf C}_{e_i}={\rm E}\left[\left(\mathbf{\Delta}_i\mathbf{\boldsymbol{\theta}}+\mathcal{I}_i\boldsymbol{\delta}_i\right)\left(\mathbf{\Delta}_i\mathbf{\boldsymbol{\theta}}+\mathcal{I}_i\boldsymbol{\delta}_i\right)^H\right]\overset{(a)}{=}{\rm E}\left[\mathbf{\Delta}_i\boldsymbol{\theta}\boldsymbol{\theta}^H\mathbf{\Delta}_i^H\right]+\mathcal{I}_i{\rm E}\left[\boldsymbol{\delta}_i\boldsymbol{\delta}_i^H\right]\overset{(b)}{=}(N\rho_i^2 + \mathcal{I}_i\rho_i'^2){\bf I}_{N_R},
\end{align}
where $(a)$ follows from the fact that the errors are uncorrelated, and $(b)$ follows from using the identity ${\rm E}\left[\mathbf{\Delta}_i\boldsymbol{\theta}\boldsymbol{\theta}^H\mathbf{\Delta}_i^H\right]=\rho_i^2{\rm Tr}\left(\boldsymbol{\theta}\boldsymbol{\theta}^H\right){\bf I}_{N_R}$.

\begin{lemma} \label{lem:NoiseRISdependence}
In Assumption I, the re-radiated signal from the ${\rm Tx}_i$ is modeled as additive Gaussian noise with variance $\zeta\sigma_{m,i}^2$ for the ${\rm Tx}_i$, where $\sigma_{m,i}^2=\mathcal{I}_i\sigma_{m_1,i}^2+N\sigma_{m_2,i}^2$ with $\sigma_{m_1,i}^2=\left(\frac{c}{4\pi f d_i}\right)^2P_i[1-\tau(f,d_i)]$ and $\sigma_{m_2,i}^2=\left(\frac{c^2}{16(\pi f)^2 }\frac{1}{d_\alpha d_{\gamma_i}}\right)^2P_i[1-\tau(f,d_\alpha)\tau(f,d_{\gamma_i})]$.
\begin{IEEEproof}
See Appendix \ref{sec:Lem1Proof}.
\end{IEEEproof}
\end{lemma}
We define $\sigma_{m_1}^2=\sum\limits_{i=0}^{N_I}\mathcal{I}_i\sigma_{m_1,i}^2$ and $\sigma_{m_2}^2=\sum\limits_{i=0}^{N_I}\sigma_{m_2,i}^2$ using Lemma \ref{lem:NoiseRISdependence} for convenient notation. This completes the SINR \eqref{eq:SINRexpression} description. 

\section{Robust Optimization of Receive Beamformer and RIS Configuration Vector} \label{sec:OpProblems}
In this section, we jointly optimize the receive beamforming weights and the RIS configuration vector. We assume that ${\rm Rx}_0$ knows the estimation error variances in the robust case. As the exact channel capacity is not computable, we set the objective function as a well-known lower bound of the channel capacity \eqref{eq:ergocap} under imperfect CSI. Maximizing a lower bound is useful in its own right as it effectively maximizes the channel capacity. However, maximizing $\log_2\left(1+\gamma(\mathbf{u},\boldsymbol{\varphi})\right)$ is equivalent to maximizing $\gamma(\mathbf{u},\boldsymbol{\varphi})$ due to the monotonically increasing nature of logarithm function. The new objective function \eqref{eq:SINRexpression} is non-convex due to the coupling between the two sets of variables. The non-convex constraint of unit modulus makes the problem more difficult. We employ the BCD method to solve this optimization problem by separating it into two sub-problems, one for each set of variables, and solving each sub-problem in an alternating manner, to obtain an efficient solution. With a fixed RIS phase vector, the receive beamformer sub-problem can be conveniently posed as a maximization of a Rayleigh quotient problem, and hence has a simple analytical solution. However, the RIS sub-problem does not have an analytical solution. We propose three methods of different complexities to achieve suboptimal solutions: a) SDP with Gaussian randomization, b) SA, and c) GD approach. Finally, we show the convergence and compare the complexities of the proposed algorithms.

\subsection{Problem Formulation}
Our objective is to maximize the effective SINR of ${\rm Tx}_0$ given the estimated channel such that the norm of receive beamforming vector is unity, and the RIS elements follow unit modulus constraint. The transmit power $P_0$ of the user is set as the maximum allowable power to maximize the objective. 
Now, the optimization problem can be formulated as:
\begin{subequations}
\label{eq:sinropt}
\begin{align}
\max_{\mathbf{u},\boldsymbol{\varphi}} \quad & \gamma(\mathbf{u},\boldsymbol{\varphi}) \label{eq:sinroptobj}\\
\textrm{s.t.} \quad & \|\mathbf{u}\|_2^2 = 1,
\\
& 0 \leq \mathbf{\varphi}_n < 2\pi, \quad \forall n=1, 2, \hdots, N.
\label{eq:sinroptcons}
\end{align}
\end{subequations}

The general BCD algorithm used for the optimization here is shown in Algorithm \ref{algo:Algo1}. Note that under imperfect CSI, initializing the $\{\rho_i,\rho'_i\}$ to zero results in the non-robust counterpart of this algorithm. The non-robust algorithm treats the estimated CSI as the perfect instantaneous CSI and is used as benchmark for the performance of the robust algorithms.
\subsection{Receive Beamformer Optimization}
As noted before, for a given $\mathbf{\boldsymbol{\theta}}$, this sub-problem can be expressed as an unconstrained maximization of a Rayleigh quotient:
\begin{align}
\max_{\mathbf{u}} \quad & \frac{\mathbf{u}^H\mathbf{B}_0\mathbf{u}}{\mathbf{u}^H\left(\sum\limits_{i=1}^{N_I} {P_i}\mathbf{B}_i +(\rho_{\rm total}+\sigma_w^2+\zeta\sigma_{m}^2)\mathbf{I}_{N_R}\right)\mathbf{u}}, \notag\\
\textrm{s.t.} \quad & \|\mathbf{u}\|_2^2 = 1,\label{eq:RxBfOp}
\end{align}
where $\rho_{\rm total}=\sum\limits_{j=0}^{N_I} {P_j}(N\rho_j^2+\mathcal{I}_j\rho_j'^2)$, $\mathbf{B}_i=\hat{\mathbf{H}}_i\mathbf{\Psi}\hat{\mathbf{H}}_i^H$, and $\mathbf{\Psi}= \boldsymbol{\theta}_0\boldsymbol{\theta}^H_0$. The normalized analytical solution for the given problem is given by \cite{tse}:
\begin{align}
\mathbf{u}^*=\frac{\left(\sum\limits_{i=1}^{N_I} {P_i}\mathbf{B}_i + (\rho_{\rm total}+\sigma_w^2+\zeta\sigma_{m}^2)\mathbf{I}_{N_R}\right)^{-1} \mathbf{e}_0}{\left\|\left(\sum\limits_{i=1}^{N_I} {P_i}\mathbf{B}_i + (\rho_{\rm total}+\sigma_w^2+\zeta\sigma_{m}^2)\mathbf{I}_{N_R}\right)^{-1} \mathbf{e}_0\right\|}, \label{eq:RxBfOptimum}
\end{align}
where $\mathbf{e}_0=\hat{\mathbf{H}}_0\boldsymbol{\theta}_0$ is the dominant eigenvector of ${\bf B}_0$ and the solution is normalized to ensure a unit norm vector.

\subsection{RIS Optimization through SDP}

We approach this sub-problem through SDP because optimality can be ensured if we can express it as a convex optimization problem. To that end, we express \eqref{eq:SINRexpression} for a given receive beamformer $\mathbf{u}$ as:
\begin{align}
    \gamma(\mathbf{u},\mathbf{\Psi})=\frac{\boldsymbol{\theta}_0^H{\bf G}_0\boldsymbol{\theta}_0}{\boldsymbol{\theta}_0^H{\bf M}\boldsymbol{\theta}_0+\alpha}\overset{(a)}{=}\frac{\mathrm{Tr}({\mathbf{\Psi}{\mathbf{G}_{0}}})
}{\mathrm{Tr}(\mathbf{\Psi}\mathbf{\mathbf{M}})+\alpha},
\end{align}
where ${\bf G}_i={P_i}\hat{\mathbf{H}}_i^H\mathbf{U}\hat{\mathbf{H}}_i$, $\mathbf{M}=\sum\limits_{i=1}^{N_I}\mathbf{G}_i+\frac{\rho_{\rm total}}{N}\mathbf{I}_{N+1}+\zeta{\sigma_{m_2}^2}\mathbf{I}_{N+1}$, $\alpha=\sigma_w^2+\zeta(\sigma_{m_1}^2-\sigma_{m_2}^2)-\frac{\rho_{\rm total}}{N}$, and (a) follows from utilizing trace operator and rearranging the terms. This reformulation imposes a positive semidefiniteness and a rank constraint on $\mathbf{\Psi}$. Unlike the first constraint, the second rank constraint that  $\textrm{rank}(\mathbf{\Psi})$ should be unity is non-convex. Relaxing this constraint is termed as SDR. The sub-problem with SDR is expressed as follows:
\begin{align}
\max_{\mathbf{\Psi}} \quad & \frac{\mathrm{Tr}({\mathbf{\Psi}{\mathbf{G}_{0}}})
}{\mathrm{Tr}(\mathbf{\Psi}\mathbf{\mathbf{M}})+\alpha},
\nonumber\\
\textrm{s.t.} \quad & 
\mathbf{\Psi} \succeq 0,
\nonumber\\
\quad & \left[\mathbf{\Psi}\right]_{l,l} = 1, \quad \forall l=1,2,\hdots,N+1. \label{eq:RISmain}
\end{align}

Following \cite{Palomar_Eldar_2009}, we introduce an auxiliary variable $b \geq 0$ to transform \eqref{eq:RISmain} to an epigraph form:
\begin{align}
\gamma^*=\max_{\mathbf{\Psi},b\geq 0} \quad & b
\nonumber\\
\textrm{s.t.} \quad & \mathrm{Tr}({\mathbf{\Psi}{\mathbf{G}_{0}}}) \geq b\mathrm{Tr}(\mathbf{\Psi}\mathbf{\mathbf{M}})+b\alpha,
\nonumber \\
\quad & \mathbf{\Psi} \succeq 0,
\nonumber\\
\quad & \left[\mathbf{\Psi}\right]_{l,l} = 1, \quad \forall l=1,2,\hdots,N+1.
\end{align}

With $b \geq 0$, the inherent feasibility problem is:
\begin{align}
\mathrm{Find} \quad & \mathbf{\Psi}
\nonumber\\
\textrm{s.t.} \quad & \mathrm{Tr}({\mathbf{\Psi}{\mathbf{G}_{0}}}) \geq b\mathrm{Tr}(\mathbf{\Psi}\mathbf{\mathbf{M}})+b\alpha,
\nonumber \\
\quad & \mathbf{\Psi} \succeq 0,
\nonumber\\
\quad & \left[\mathbf{\Psi}\right]_{l,l} = 1,\quad \forall l=1,2,\hdots,N+1. \label{eq:RISOptProblem}
\end{align}

Note that if the above problem is feasible then $\gamma(\mathbf{u},\mathbf{\Psi})^*\geq b$, while the opposite condition $\gamma(\mathbf{u},\mathbf{\Psi})^*\leq b$ holds when the above problem is infeasible. So, using bisection for \eqref{eq:RISOptProblem} provides a good solution for $\bf \Psi$. The problem \eqref{eq:RISOptProblem} can be solved by any standard convex optimization package like CVX \cite{cvx,gb08}. However, this solution will generally not be a rank-one solution. We extract a rank-one solution by Gaussian randomization \cite{Palomar_Eldar_2009}, as discussed next. The unit circle projection (the division of each entry of the vector by its absolute value) of the rank-one solution provides $\mathbf{\boldsymbol{\theta}}^*$.

\begin{algorithm}
\SetAlgoLined
\KwInput{$\mathbf{G}_{0},\mathbf{M},\alpha,G,\mathbf{u}_{i+1}~\forall i$}
\KwOutput{$\bar{\mathbf{\boldsymbol{\theta}}}_{{i+1}}$} 
  Obtain $\mathbf{\Psi}_{i+1}$ by using bisection on (\ref{eq:RISOptProblem}).\\
  Generate $G$ feasible solutions for $\bar{\mathbf{\boldsymbol{\theta}}}_{0_{i+1}}$ through Gaussian randomization \cite{Palomar_Eldar_2009}.\\
  Choose the solution with unit circle projection $\bar{\mathbf{\boldsymbol{\theta}}}_{{i+1}}$ that provides the highest $\gamma_{i+1}$ through (\ref{eq:SINRexpression}).\\

  \caption{BCD-SDR} \label{algo:Algo2}
\end{algorithm}
\subsubsection*{Gaussian Randomization}

The Gaussian randomization scheme \cite{gauss} entails generation of a random vector ${\bf z}\sim\mathcal{CN}({\bf 0}_{N+1},{\bf \vartheta})$ where ${\bf \vartheta}$ is the Cholesky decomposition of ${\bf \Psi}$ to extract a rank-one solution. The rank-one candidate for one such vector is $\left[\frac{{\bf z}_N^T}{\|\mathbf{z}_N\|}~1\right]^T$, where $\mathbf{z}_N$ is the vector consisting of the first $N$ elements of $\bf z$. A number of such candidates are generated to choose the one that is feasible and provides the largest objective value among them. The approximation accuracy of such a scheme in different scenarios is well-investigated by \cite{gauss}. The worst-case approximation accuracy was proved to be reasonable for a finite number of generations. However, the complexity of this problem is prohibitive for RISs with large number of elements \cite{lowc}. This encourages us to explore some low-complexity approaches in the subsequent subsections.

\subsection{SA Solution to the RIS Sub-problem}
A low-complexity approach to the RIS sub-problem is to maximize the received signal strength as opposed to SINR due to the existence of a closed-form solution \cite{wu2019beamforming}. As this approach aligns the phases of reflected signal with the phase of the direct signal, we denote this approach as SA and is expressed as follows:
\begin{align}
    {\bf \boldsymbol{\theta}}_i=e^{-j\left(\arg({\bf u}^H\hat{{\bf Z}}_0)_i-\arg({\bf u}^H\hat{{\bf h}}_{{\rm RT}_0})\right)}.\label{eq:heusol}
\end{align}

This can incur some performance loss from the global optimum solution. Since most RIS phase-shift optimization problems are NP-hard, a global optimum is difficult to obtain \cite{opti}. However, it has been demonstrated that the local optimal solutions can improve performance significantly \cite{sp,ee}. In that case, the performance loss can be characterized by finding good stationary points. In \cite{sp}, a stationary point of a general SINR is found with respect to the phase-shift of one element without considering direct links to facilitate the element-wise BCD algorithm. Meanwhile, the authors of \cite{activeRIS} obtain a stationary point of an SINR expression in an $N$-element active RIS-aided network that does not consider interference. This was possible as the active RIS elements do not have the unit modulus constraint of passive RIS. However, a similar analysis has not been done for the passive RIS. To fill this gap, we provide an analysis next for a one-element passive RIS with only one interferer while considering direct links.
\subsubsection*{One-element RIS Sub-problem} A generic SINR expression for an one-element RIS with only one interferer can be readily written as:
\begin{align}
    \gamma_1=\frac{P_0|a_1\theta_1+h|^2}{P_1|b_1\theta_1+g|^2+c}=\frac{L'+M'\cos(s+x)}{N'+P'\cos(t+x)}, \label{twoSINR}
\end{align}
where $a_1,b_1,\theta_1,\text{ and }c$ denote the reflected signal channel, reflected interferer channel, RIS phase shifts and noise. The direct signal and interferer channels are denoted by $h$ and $g$, respectively. However, we will use the second form to derive the stationary points where $L'=P_0(|a_1|^2+|h|^2), M'=2P_0|a_1||h|, N'=P_1(|b_1|^2+|g|^2)+c, P'=2P_1|b_1||g|, s=\angle{a_1}-\angle{h}, t=\angle{b_1}-\angle{g}$, and $x=\angle\theta_1 $.
\begin{theorem}
The SINR for one-element RIS takes one of the two values at the stationary points given by:
\begin{align}
    \gamma_1^*=\frac{L'}{N'}-\frac{1}{N'}\frac{C'}{P'(L'P'-M'N'\cos(s-t))\pm N'\sqrt{C'-(M'P'\sin(s-t))^2}}, \label{eq:SPsolution}
\end{align}
where $C'=(L'P')^2+(M'N')^2-2L'M'N'P'\cos(s-t)$.
\label{lem:SPsolution}
\end{theorem}
\begin{IEEEproof}
See Appendix \ref{sec:TheoProof}.
\end{IEEEproof}

\begin{remark}
In the simple setting without interference or $P_1\to0$, the SINR $\gamma_1^* \to \frac{L'+M'}{N'} $. This is a result of choosing $\theta_1$ in a way to align the phase of $a_1\theta_1$ with $h$, which is the SA method. In absence of interference, the optimal solution is the SA solution.
\label{rem:remSA}
\end{remark}

\begin{cor}
In a simple setting where $|a_1|, |b_1|, |h|, P_1, P_0$ are unity and $|g|=k$, the SINR for the SA solution is $g_1(k)=\frac{16k^2\sin^2(s-t)}{(k^2+2k\cos(s-t)+1+c)((k+1)^2+c)((k-1)^2+c)}$ below the higher stationary point and $g_2(k)=\frac{4}{k^2+2k\cos(s-t)+1+c}$ above the lower stationary point. \label{cor:gap}
\end{cor}
\begin{IEEEproof}
The SA solution in this case is $x=-s$. The resulting SINR is denoted by $\gamma_2^h=\frac{L'+M'}{N'+P'\cos(s-t)}$. Both $g_1(k)$ and $g_2(k)$ can be calculated easily by evaluating $|\gamma_1^*-\gamma_2^h|$.
\end{IEEEproof}
\begin{remark}
Note that $g_1(k)$ and $g_2(k)$ both tend to zero when $k \to \infty$. In other words, the stationary points converge to the SA solution when the interferer direct link is too powerful. In a practical scenario, where the interferer direct link is as powerful as the reflected link or $k \to 1$, $g_1(k) \to \frac{16\sin^2(s-t)}{(4c+c^2)(2\cos(s-t)+2+c)}$ implying the sub-optimality of the SA solution. \label{rem:highid}
\end{remark}
These results confirm the need for a low-complexity algorithm that finds a better solution than the SA, which inspires our next approach.
\subsection{Gradient Descent Approach to the RIS Sub-problem}
Achieving a better solution than the SA method while retaining its low-complexity benefit requires a different approach. As noted in the previous subsection, local optimal solutions can provide significant performance improvement. Gradient descent is a natural choice for such a solution because it tries to converge to a local minimum from the initial point in the chosen descent direction. Generally, the chosen descent direction is the steepest one or the negative gradient of the objective function. As shown next, this can be calculated from an alternate formulation of our RIS sub-problem in an unconstrained manner with respect to $\boldsymbol{\varphi}=[\varphi_1\,\hdots\,\varphi_n]^T$:
\begin{align}
\min_{\boldsymbol{\varphi}} \quad &  -\frac{\mathbf{\boldsymbol{\theta}}^H{\mathbf{R}_{0}}
\mathbf{\boldsymbol{\theta}}+2{\rm Re}(\bf{c}_0{\bf\boldsymbol{\theta}})}{\mathbf{\boldsymbol{\theta}}^H{\bf K} \mathbf{\boldsymbol{\theta}}+2{\rm Re}(\bf{z}{\bf\boldsymbol{\theta}})}, \label{eq:GDRIS}
\end{align}
where 
$
{\bf R}_i={P_i}\left(\hat{\mathbf{Z}}_i^H\mathbf{U}\hat{\mathbf{Z}}_i+\frac{|\mathcal{I}_i{\bf u}^H\hat{\bf h}_{{\rm RT}_i}|^2}{N}\mathbf{I}_{N}\right),{\bf K}=\sum\limits_{i=1}^{N_I}{\bf R}_i+\left(\frac{\rho_{\rm total}+\sigma_w^2+\zeta\sigma_{m_1}^2}{N}+\zeta\sigma_{m_2}^2\right)\mathbf{I}_{N}$, ${\bf c}_i={P_i}\mathcal{I}_i\hat{\bf h}_{{\rm RT}_i}^H{\bf u}{\bf u}^H\hat{\mathbf{Z}}_i$, and ${\bf z}=\sum\limits_{i=1}^{N_I} {\bf c}_i.
$ The gradient can be calculated as follows:
\begin{align}
    &{\bf \nabla}_{\boldsymbol{\varphi}}\left(\boldsymbol{\varphi}\right)=2\textrm{Re}\left\{\frac{({\bf R}_0^*{\bf \boldsymbol{\theta}}^*+{\bf c}_0^T)\odot(-j{\bf\boldsymbol{\theta}})}{\mathbf{\boldsymbol{\theta}}^H{\bf K}\mathbf{\boldsymbol{\theta}}+2{\rm Re}(\bf{z}{\bf\boldsymbol{\theta}})}\right\}\notag\\&+2\textrm{Re}\left\{\frac{\left(\mathbf{\boldsymbol{\theta}}^H{\bf R}_0\mathbf{\boldsymbol{\theta}}+2{\rm Re}(\bf{c}_0{\bf\boldsymbol{\theta}})\right)\cdot({\bf K}^*{\bf \boldsymbol{\theta}}^*+{\bf z}^T)\odot(j{\bf\boldsymbol{\theta}})}{\left(\mathbf{\boldsymbol{\theta}}^H{\bf K}\mathbf{\boldsymbol{\theta}}+2{\rm Re}(\bf{z}{\bf\boldsymbol{\theta}})\right)^2}\right\}.\label{eq:gradf}
\end{align}
\begin{remark}
By substituting $\{{\bf C}_i\}$ as zero vectors, we obtain the same objective function and the gradient as \cite{lowc}.
\end{remark}
Armed with the analytical expression of the gradient, a simple GD algorithm works through the simple update rule:
\begin{align}
    \boldsymbol{\varphi}^{(t+1)}=\boldsymbol{\varphi}^{(t)}-\beta^{(t)}{\boldsymbol {\nabla}}_{\boldsymbol{\varphi}}\left(\boldsymbol{\varphi}^{(t)}\right),
\end{align}
where $\boldsymbol{\varphi}^{(t)}$ is the RIS phase vector at $t$-th iteration, $\beta^{(t)}$ is the step-size and ${\boldsymbol{ \nabla}}_{\boldsymbol{\varphi}}\left(\boldsymbol{\varphi}^{(t)}\right)$ is the gradient. In practice, convergence time is dependent on choosing a good step-size. A step-size that is either too large or too small can result in slow convergence by either oscillating or moving too slowly in the descent direction. A practical alternative is to implement backtracking line searches. These line searches start with an initial step-size and keep diminishing them in a loop. The largest step-size in that sequence is chosen that ensures a sufficient decrease in the descent direction. Moreover, \cite{vanish} shows that GD with diminishing step sizes almost always avoids saddle points under random initialization. Saddle points are the critical points that are neither local minima nor local maxima. It is also shown in \cite{escape} that saddle points can slow down GD with constant step size considerably to the extent of needing exponential time to escape, even with reasonable random initialization schemes. The objective function is highly non-convex and expected to have multiple saddle points, so being able to avoid saddle points is a desirable property. With these motivating factors, we choose Armijo-Goldstein (AG) line search \cite{agline}. This strategy ensures that $\beta^{(t)}$ satisfies
\begin{align}
    &-\gamma\left(\mathbf{u}^{(t)},\boldsymbol{\varphi}^{(t)}-\beta^{(t)}{\boldsymbol  {\nabla}}_{\boldsymbol{\varphi}}\left(\boldsymbol{\varphi}^{(t)}\right)\right)\leq -\gamma\left(\mathbf{u}^{(t)},\boldsymbol{\varphi}^{(t)}\right)-\varepsilon \beta^{(t)}\|{\boldsymbol {\nabla}}_{\boldsymbol{\varphi}}\left(\boldsymbol{\varphi}^{(t)}\right)\|_2^2,\label{eq:armijo}
\end{align}
where $0<\varepsilon<1$ is a constant. If the condition \eqref{eq:armijo} is violated, the step-size is decreased by a factor of $0<\varrho<1$ starting from a larger initial step-size. Complete details of the GD approach are shown in Algorithm \ref{algo:Algo3}. Note that, our proposed algorithm uses the SA solution as the initial point. With all the sub-problem solutions, the general BCD framework is demonstrated in Algorithm \ref{algo:Algo1}.

\begin{algorithm}
\SetAlgoLined
\KwInput{$\mathbf{L}_{0}, \mathbf{Z},\varrho, \varepsilon, \epsilon_{th}, \beta_0, \mathbf{u}_{i+1}~\forall i$}
\KwOutput{$\bar{\mathbf{\boldsymbol{\theta}}}_{{i+1}}$}
Initialize $t=1$, $\delta_{GD}=1$, and $\boldsymbol{\varphi}^{(t)}=-\arg({\bf u}_{i+1}^H{\bf H}_0)_i$.\\
\While{$\delta_{GD} \leq \epsilon_{th}$\\}
{
Initialize $\beta^{(1)}=\beta_0$.\\
Calculate ${\boldsymbol {\nabla}}_{\boldsymbol{\varphi}}\left(\boldsymbol{\varphi}^{(t-1)}\right)$ from \eqref{eq:gradf}.\\
\While{$-\gamma_{\rm LB}\left(\mathbf{u}^{(t)},\boldsymbol{\varphi}^{(t)}-\beta^{(t)}{\boldsymbol {\nabla}}_{\boldsymbol{\varphi}}\left(\boldsymbol{\varphi}^{(t)}\right)\right)\geq -\gamma_{\rm LB}\left(\mathbf{u}^{(t)},\boldsymbol{\varphi}^{(t)}\right)-\varepsilon \beta^{(t)}\|{\boldsymbol {\nabla}}_{\boldsymbol{\varphi}}\left(\boldsymbol{\varphi}^{(t)}\right)\|_2^2$\\}{$\beta^{(t)}=\varrho\beta^{(t)}$.}
$\delta_{GD}=\beta^{(t)}\|{\boldsymbol {\nabla}}_{\boldsymbol{\varphi}}\left(\boldsymbol{\varphi}^{(t)}\right)\|_2^2$.\\
$\boldsymbol{\varphi}^{(t+1)}=\boldsymbol{\varphi}^{(t)}-\beta^{(t)}{\boldsymbol {\nabla}}_{\boldsymbol{\varphi}}\left(\boldsymbol{\varphi}^{(t)}\right).$\\
$t=t+1.$
}
$\bar{\mathbf{\boldsymbol{\theta}}}_{{i+1}}=\exp(-j\boldsymbol{\varphi}^{(t)})$.
  \caption{BCD-GD} \label{algo:Algo3}
\end{algorithm}
\begin{algorithm}
\SetAlgoLined
\KwInput{$\mathbf{H}_{i},\zeta,\epsilon~\forall i$}
\KwOutput{$\mathbf{\boldsymbol{\theta}}^*,\mathbf{u}^*$} 
Initialize $\mathbf{\boldsymbol{\theta}}$ with a random vector, $i=0$, $\gamma_0=0$, and $\Delta=\epsilon+1$.\\
 \While{$\Delta>\epsilon$}{
  Obtain $\mathbf{u}_{i+1}$ from (\ref{eq:RxBfOptimum}).\\
  Obtain $\mathbf{\bar{\boldsymbol{\theta}}}_{{i+1}}$ from Algorithm \ref{algo:Algo2}, \ref{algo:Algo3} or \eqref{eq:heusol}.\\
 \If{$\gamma_{i+1}\leq\gamma_i$}{$\bar{\mathbf{\boldsymbol{\theta}}}_{{i+1}}=\bar{\mathbf{\boldsymbol{\theta}}}_{{i}}$.\\
   }{}
   {
  
  }
  Evaluate $\Delta=|\gamma_{i+1}-\gamma_{i}|/\gamma_{i}$.\\
  $i=i+1$.
 }
 $\mathbf{{\boldsymbol{\theta}}}^*=\mathbf{\bar{\boldsymbol{\theta}}}_{{i-1}}$.
 \caption{Joint Optimization by BCD} \label{algo:Algo1}
\end{algorithm}
\subsection{Convergence and Complexity Discussion}

The BCD algorithm presented in Algorithm \ref{algo:Algo1} ensures non-decreasing objective values after each iteration. As the objective function is also upper bounded by some value, Algorithm \ref{algo:Algo1} converges regardless of the approach taken to solve the RIS sub-problem. However, all the approaches are of varying complexities. In this subsection, worst-time complexities are derived in the big-O notation for the algorithms when tractable and run-times are compared numerically. First, for each iteration of the BCD algorithm, we will calculate the time complexity for each sub-problem.

\subsubsection{Beam-former Optimization}
The total complexity of the calculation of each $\mathbf{B}_{i}$ for all $i=0,2\hdots,N_I$ is $O(N_R^2 N (N_I+1))$. The multiplication with $\mathbf{e}_{0}$, the inverse, and the norm operation have time complexities of $O(N_R^2),O(N_R^3)$, and $O(N_R)$, respectively. So, the total time complexity is $O((N_R^2 N (N_I+1)+N_R^3+N_R^2+N_R)=O(N_R^2 N (N_I+1)+N_R^3)$.

\subsubsection{SDR Algorithm}
First, the calculation of $N_I+1$ $\{\mathbf{G}_i\}$ has a time complexity of $O(((N+1)N_R+(N+1)^2)(N_I+1))=O(N(N_I+1)(N+N_R))$. The feasibility problem (\ref{eq:RISOptProblem}) is a classic semi-definite programming problem and is solved by interior-point method. The worst time complexity is $O((N+1)^{4.5})$ \cite{timecomplexity1,tc2}. This computation happens every iteration of the bisection algorithm. The number of iterations needed is  $\log_2\left(\frac{\epsilon_0}{\epsilon_1}\right)$, where $\epsilon_0$ and $\epsilon_1$ denote the upper bound and the tolerance for the bisection problem, respectively \cite{bisectionoptimal}. So, the worst-case time complexity of the feasibility problem (\ref{eq:RISOptProblem}) with the bisection procedure and the matrix multiplications is $O\left(N(N_I+1)(N+N_R)+(N+1)^{4.5}\log_2\left(\frac{\epsilon_0}{\epsilon_1}\right)\right)=O\left(N(N_I+1)(N+N_R)+N^{4.5}\log_2\left(\frac{\epsilon_0}{\epsilon_1}\right)\right)$. For the Gaussian randomization procedure, we need to do a Cholesky factorization to create a normal random variable with the covariance matrix $\mathbf{\Psi}^*$ as per \cite{Palomar_Eldar_2009}. The time complexity of this operation is $O((N+1)^3)=O(N^3)$. Next, the creation of $G$ number of random samples requires a time complexity of $O(G(N+1)^2) = O(GN^2)$. Finally adding both of them together and ignoring lower order terms, the SDR algorithm has a time complexity of $O\left(N^{4.5}\log_2\left(\frac{\epsilon_0}{\epsilon_1}\right)+N^3+N(N_I+1)(N+N_R)+GN^2\right)$.

\subsubsection{SA Solution}
The SA solution has a time complexity of $O(N_R^2N)$ resulting from the multiplication between ${\bf u}^H$ and ${\bf H}_0$.

\subsubsection{GD Algorithm}
The complexity of this method is dominated by the gradient and initial point calculation. Assuming $I_1$ iterations of the GD algorithm, one iteration of gradient calculation requires pre-calculating ${\bf R}_i$ and ${\bf c}_i$ for all $i=0,1,\hdots,N_I$ along with the calculation of the gradient expression. The former incurs a worst-time complexity of $O((N_I+1)(N^2+2N_RN))$ while the latter has a time complexity of $O(4N^2)$. Considering the initial point calculation, the total time complexity is $O(I_1((N_I+1)(N^2+2N_RN)+4N^2)+N_R^2N)$.
Now the run-times for one iteration at $(N_R=100)$ and $N=100$ are compared in Fig. \ref{fig:runtime} and compiled in Table \ref{tab:compare} along with the complexities. The algorithm with the SA sub-problem solution is denoted by BCD-SA.

\begin{table}
\begin{center}
\begin{tabular}{|c|p{80mm}|c|}
\hline
RIS sub-problem & Complexity & Runtime in seconds \\ \hline
BCD-SDR & $O(N^{4.5}\log_2\left(\frac{\epsilon_0}{\epsilon_1}\right)+N^3+N(N_I+1)(N+N_R)+GN^2)$ & 129.4 \\
\hline
BCD-SA & $O(N_R^2N)$ & 0.012 \\
\hline
BCD-GD & $O(I_1((N_I+1)(N^2+2N_RN)+4N^2)+N_R^2N)$ & 0.005 \\
\hline
\end{tabular}
\end{center}
\caption{\label{tab:compare}Complexity Comparison.}
\end{table}

\section{Numerical Results} \label{sec:NumResults}
In this section, we provide numerical results to quantify the impact of molecular re-radiation, RIS elements, Rx antennas, interferers, and channel estimation error on key performance metrics of an RIS-aided wireless network inspired by a 3D indoor xR setup. We consider a spherical coordinate system of $(r,\varphi,\theta)$ where $r$ denotes the distance in meters from the origin, and the angles are defined in \eqref{eq:afactor}. In the simulation setup, one corner of the ${\rm Rx}_0$ arranged in a square URA pattern with $100$ antennas is situated at the origin $(0,0^\circ,0^\circ)$ on the positive y-z plane while the location and orientation of the RIS (equipped with $100$ elements) are translated $1$m along the positive x-axis with respect to the ${\rm Rx}_0$ location. The RIS is also arranged in a square URA pattern. The ${\rm Tx}_0$ is placed at $(1,60^\circ,0^\circ)$ and transmits at $2$ W ($33$ dBm) effective isotropic radiated power (EIRP) (which is within typical operating parameters of THz systems, e.g., see \cite{power2w}) over a large bandwidth specified later. The coordinates of the interferer are $(1.5,110^\circ,0^\circ)$, and it transmits at the same power level. Note that, these distances are chosen to ensure that all the nodes operate in a far-field region. Unless otherwise specified, the system parameters for the simulation configuration are as follows: the transmission carrier frequency of $220$ GHz, bandwidth of $10$ GHz, relative humidity of $50\%$, standard atmospheric pressure of $1$ atm, and temperature of $27^\circ$C. These parameters, in conjunction, determine the value of $k(f)$. In addition, the thermal noise variance is assumed to be $-174$ dBm/Hz. For the general BCD algorithm, the parameter $\epsilon$ is set to be $10^{-6}$. The parameters $\epsilon_0, \epsilon_1$, and $G$ in the SDR sub-problem are set to be $20, 10^{-6}$, and $1000$, respectively. The parameters $\varepsilon, \epsilon_{th}, \beta_0$, and $\varrho$ in the GD sub-problem are set as $0.00005, 10^{-6}, 1$, and $0.5$, respectively. We have used throughput and uncoded symbol error rate (SER) as the performance metrics. Throughput is calculated by the expression $BC_{\rm sys}$ for the perfect CSI case and throughput results are averaged over $2000$ iterations. SER is calculated using 4-QAM modulation scheme and SER results are averaged over $10^6$ symbols. We also assume that direct signal path is completely blocked to focus on the RIS's capability. This use-case pertains to indoor THz communications, as the visibility of LoS link depends on the transmission frequency and the probability of blockage by the user's own body (self-blockage) or by other user's body (dynamic blockage) \cite{VRTHz3}.

\subsubsection*{Legend notations} In the figures, the notations `ND' and `D' denote the two cases regarding the availability of the direct links of the interferers. In particular, `ND' and `D' correspond to $P_{L_i}=0$ (\emph{no direct link}), and $P_{L_i}=1$  (\emph{direct link}) for all $i=1,2,\hdots,N_I$, respectively.  Similarly, the abbreviations `N' and `SC' denote Assumptions \ref{assumption:noise} and \ref{assumption:scatter}, respectively. Additionally, the algorithms `BCD-SA', `BCD-GD' and `BCD-SDR' are abbreviated as `SA', `GD' and `SDR'. These algorithms are compared with random phase-shift RIS and optimized receive beamformer baseline performance denoted as `RAND'. The robust and non-robust counterparts of the algorithms are denoted by `R' and `NR', respectively.

\begin{figure}
    \centering
        \begin{subfigure}[b]{0.32\textwidth}
         \centering
         \includegraphics[width=\textwidth]{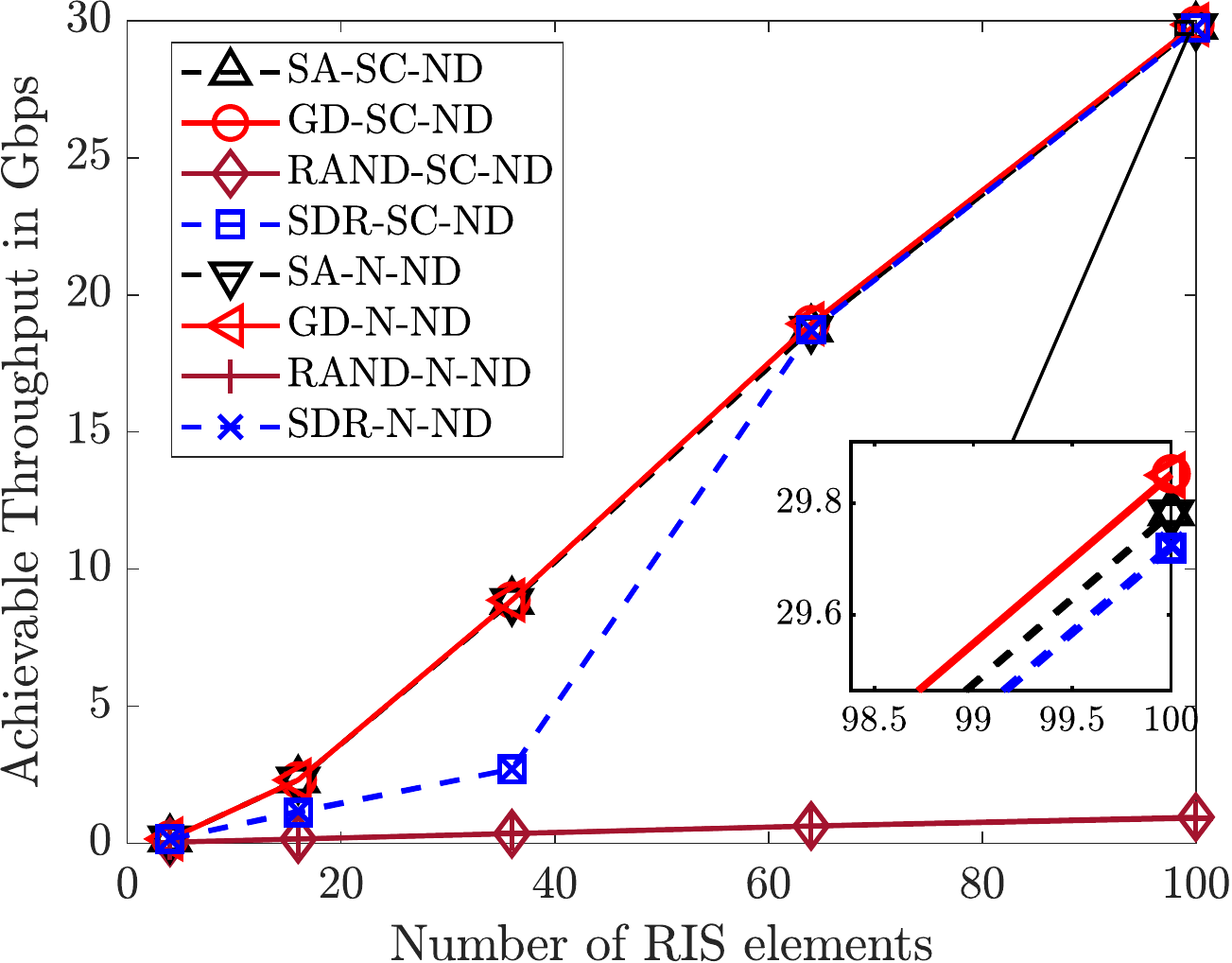}
         \caption{}
         \label{fig:risnd}
     \end{subfigure}
     \hfill
     \begin{subfigure}[b]{0.32\textwidth} 
         \centering
         \includegraphics[width=\textwidth]{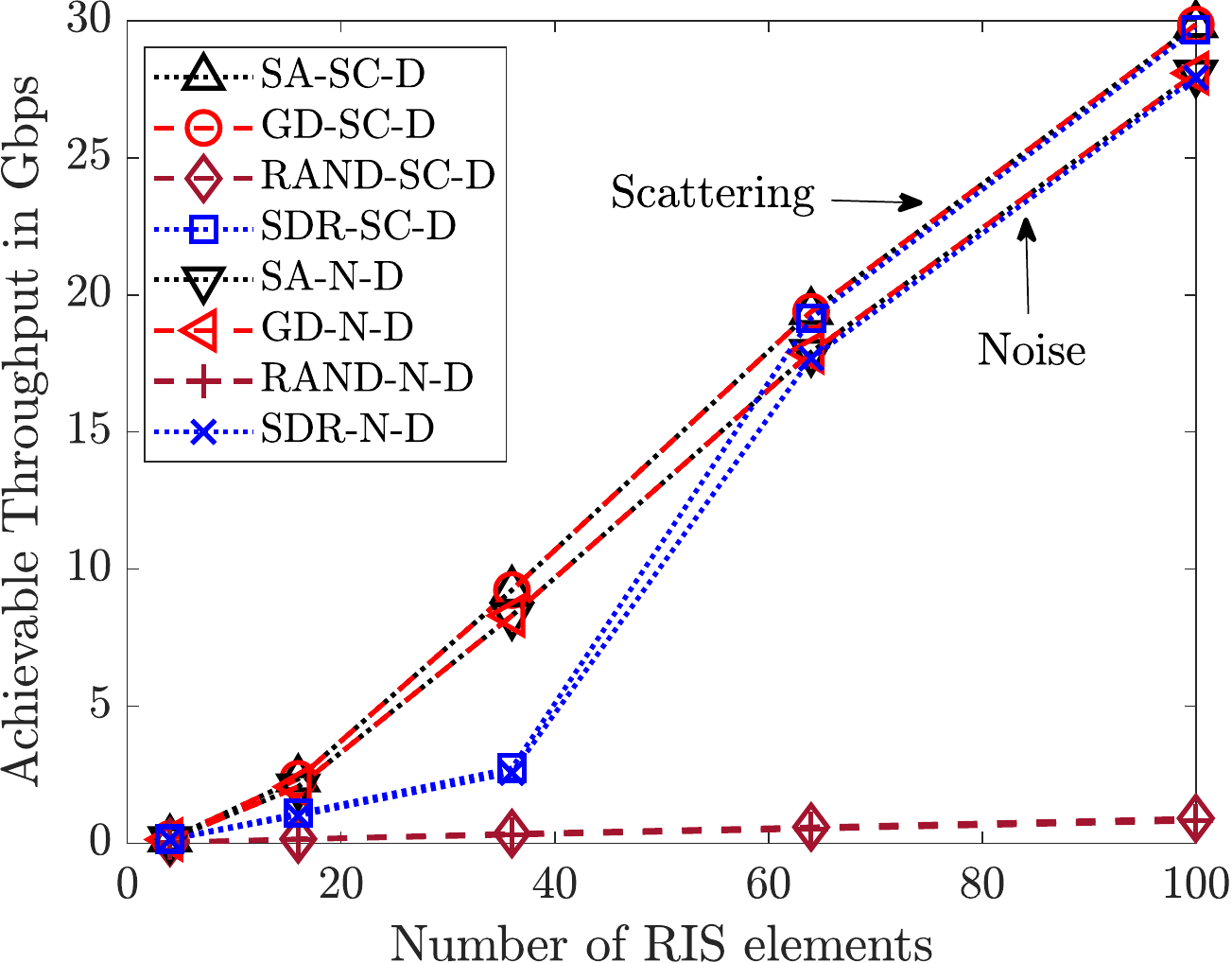}
         \caption{}
         \label{fig:risd}
     \end{subfigure}
      \begin{subfigure}[b]{0.33\textwidth} 
         \centering
         \includegraphics[width=\textwidth]{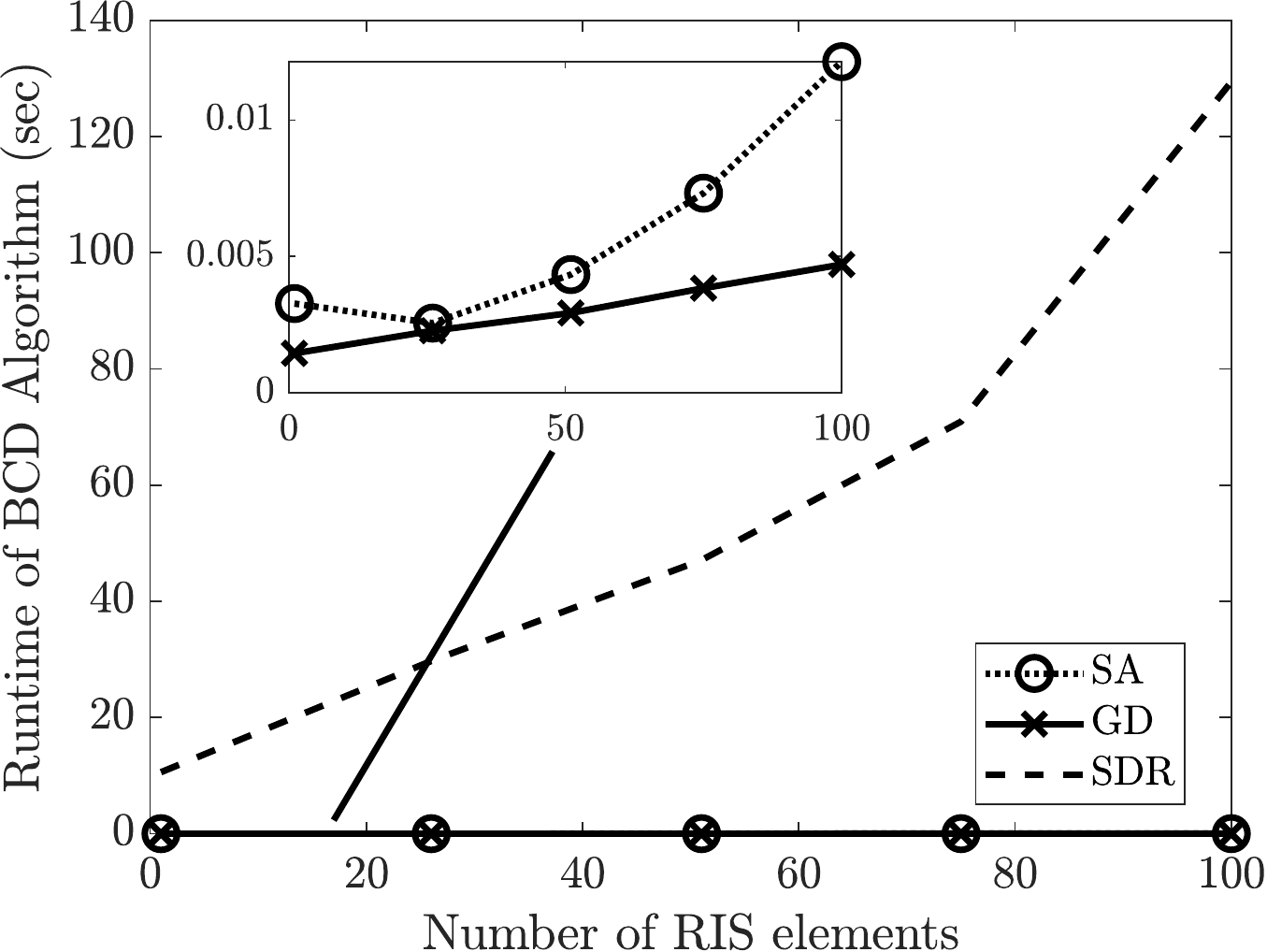}
         \caption{}
         \label{fig:runtime}
     \end{subfigure}
        \caption{ Achievable throughput with the number of RIS elements when (a) Interferer direct links do not exist, (b) Interferer direct links exist, and (c) runtime in seconds with the number of RIS elements,}
        \label{fig:RIS}
\end{figure}

\subsubsection*{Runtime comparison of different algorithms} In Fig. \ref{fig:runtime}, we plot run-times of one iteration of the three different algorithms against different numbers of RIS elements. These run-times are averaged from the simulations needed to create Fig. \ref{fig:risnd} and \ref{fig:risd} on a 3.59GHz AMD Ryzen 5 3600 6-Core PC with 16GB RAM. We observe that GD is the fastest algorithm while SDR is the slowest. Note that the run-time for the GD algorithm increases linearly unlike the SA and SDR algorithms. A possible explanation is that the GD sub-problem performs better in the BCD environment as it gives a near-optimal solution unlike a guaranteed sub-optimal solution from the SA sub-problem.

\subsubsection*{Effect of Number of RIS elements on achievable throughput} In Fig. \ref{fig:risnd} and \ref{fig:risd}, we plot achievable throughput with varying number of RIS elements. The throughput increases almost linearly with increasing number of RIS elements. The high throughput is a direct result of having a large bandwidth as the communication is relatively in the low to moderate SINR regime ($0 - 7$ dB). In this regime, the SINR is proportional to $N^2$ and translates to almost linear increase in throughput.

The optimized system provides an increase of $30$ Gbps throughput over the baseline at $N=100$. The gap between assumptions are more visible in Fig. \ref{fig:risd} where the interferer direct links are present. It is due to the high re-radiation noise power from the interferer direct path that is absent in Fig. \ref{fig:risnd}. In Fig. \ref{fig:risd}, Assumption \ref{assumption:scatter} provides almost $2$ Gbps of throughput increase from Assumption \ref{assumption:noise} when $N\in\{64,100\}$. In Fig. \ref{fig:risnd}, the SA solution is near-optimal and almost overlaps with the GD and the SDR solutions due to low interference in accordance with Remark \ref{rem:remSA}. However, in Fig. \ref{fig:risd}, as the interferer direct link is much stronger than the reflected link (as a result of our chosen path-loss model and scenario), the stationary points converge to the SA solution following Remark \ref{rem:highid}. This behavior is verified by the overlap of the GD and SA solutions. From the zoomed plot in Fig. \ref{fig:risnd}, BCD-GD performs slightly better than both BCD-SA and BCD-SDR. However, the gap between BCD-GD and BCD-SA vanishes in the Fig. \ref{fig:risd}. In summary, BCD-SA and BCD-GD both achieve similar performance to the conventional BCD-SDR under perfect CSI while being fast. Next, we investigate achievable throughput as function of the number of Rx antennas when the interferer direct links are present as the scenario `ND' does not provide much information about the assumptions.

\subsubsection*{Effect of Rx antennas and estimation of $\zeta$ on achievable throughput} 
\begin{figure}
    \centering
     \begin{subfigure}[b]{0.32\textwidth}
         \centering
         \includegraphics[width=\textwidth]{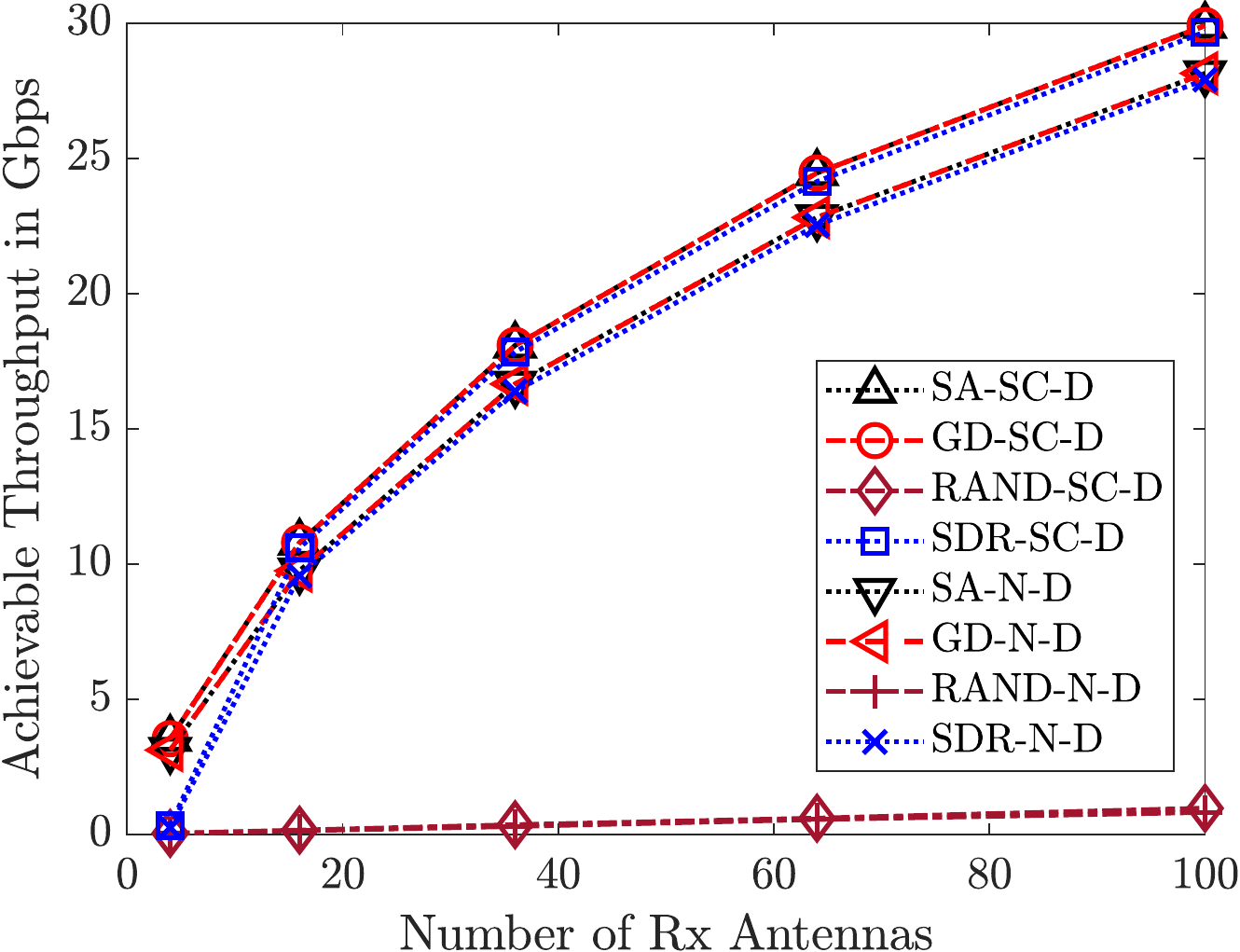}
         \caption{}
         \label{fig:rxz}
     \end{subfigure}
     \hfill
     \centering
     \begin{subfigure}[b]{0.32\textwidth} 
         \centering
         \includegraphics[width=\textwidth]{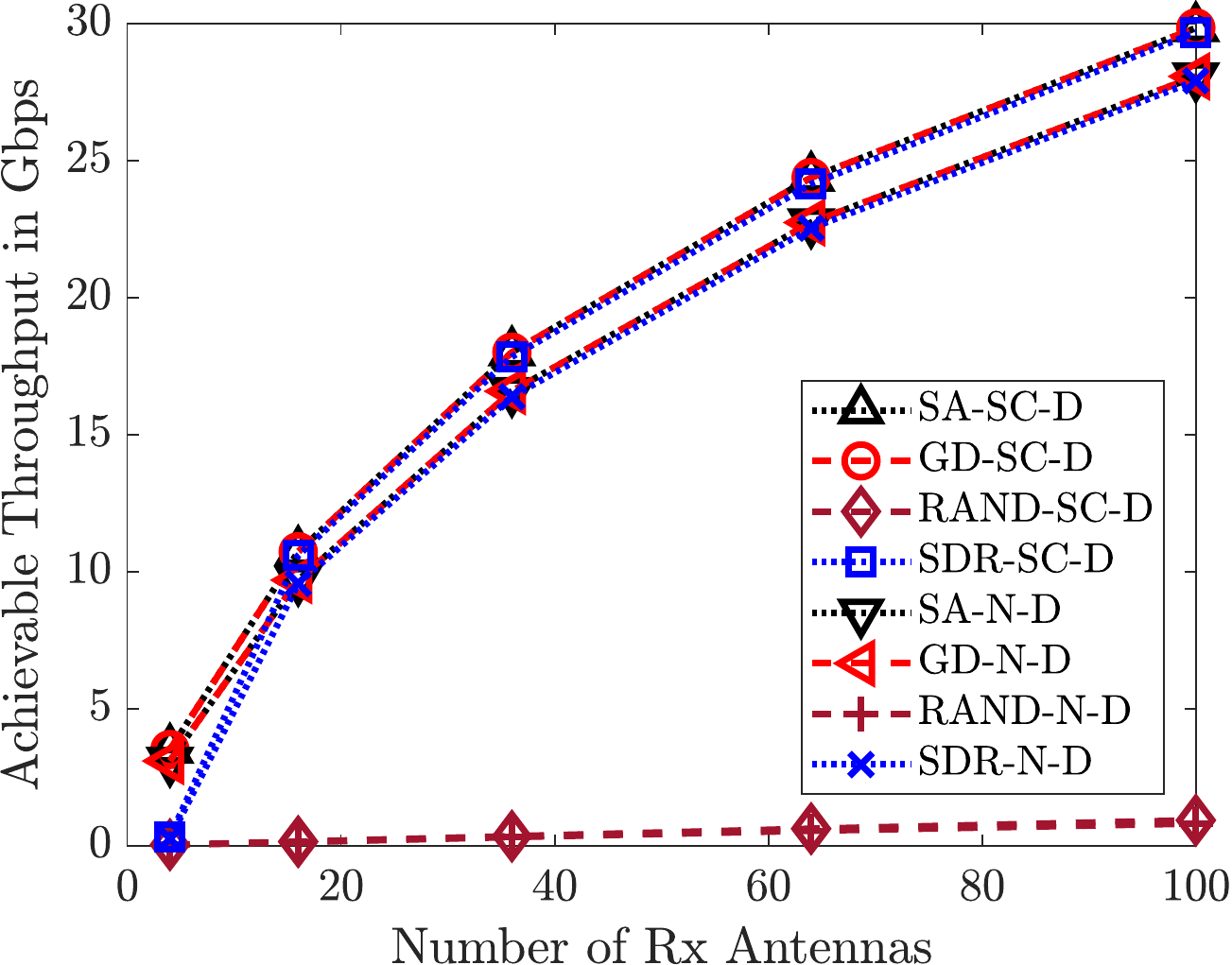}
         \caption{}
         \label{fig:rxoz}
     \end{subfigure}
          \begin{subfigure}[b]{0.33\textwidth} 
         \centering
         \includegraphics[width=\textwidth]{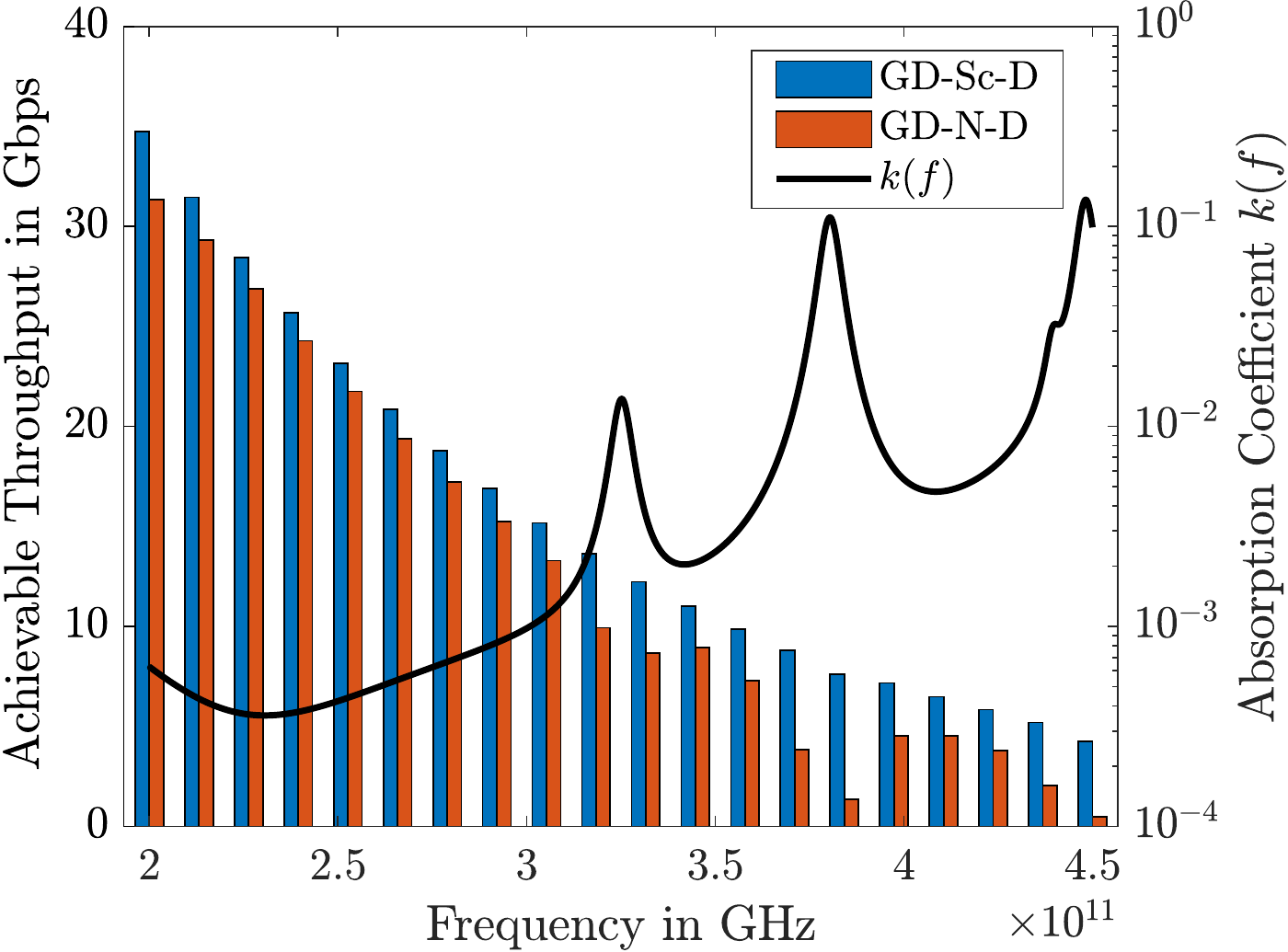}
         \caption{}
         \label{fig:freq}
     \end{subfigure}
        \caption{Achievable throughput with the number of Rx antennas when (a) perfect $\zeta$ estimation, (b) opposite $\zeta$ estimation, and (c) frequency.}
        \label{fig:Rx}
\end{figure}

In Fig. \ref{fig:rxz} and \ref{fig:rxoz}, we plot throughput versus number of Rx antennas. We investigate the case when we optimize according to Assumption \ref{assumption:noise} whereas Assumption \ref{assumption:scatter} is the reality and vice-versa. In both the cases, we assume that we have the perfect CSI for the optimization procedure. Our assumption for the optimization procedure only affects the amount of noise in the SINR expression. The cases of perfect and opposite $\zeta$ estimation are shown in Fig. \ref{fig:rxz} and \ref{fig:rxoz}, respectively. We observe from Fig. \ref{fig:Rx} that the joint optimization procedure is not very sensitive to the model assumption when perfect CSI is obtained. This is expected as the optimization procedure differs just in noise amount under different model assumptions. Furthermore, since the Rx always points to the RIS as the signal direct link is blocked, the Rx can only suppress the direct interference through its sidelobe management. This causes the SINR to scale linearly with the number of Rx antennas $N_R$. Subsequently, logarithmic increasing trend of throughput is observed with the increase of the number of Rx antennas unlike the almost linear trend observed in case of RIS elements.

\subsubsection*{Effect of CSI quality on SER} 
\begin{figure}
    \centering
     \begin{subfigure}[b]{0.32\textwidth}
         \centering
         \includegraphics[width=\textwidth]{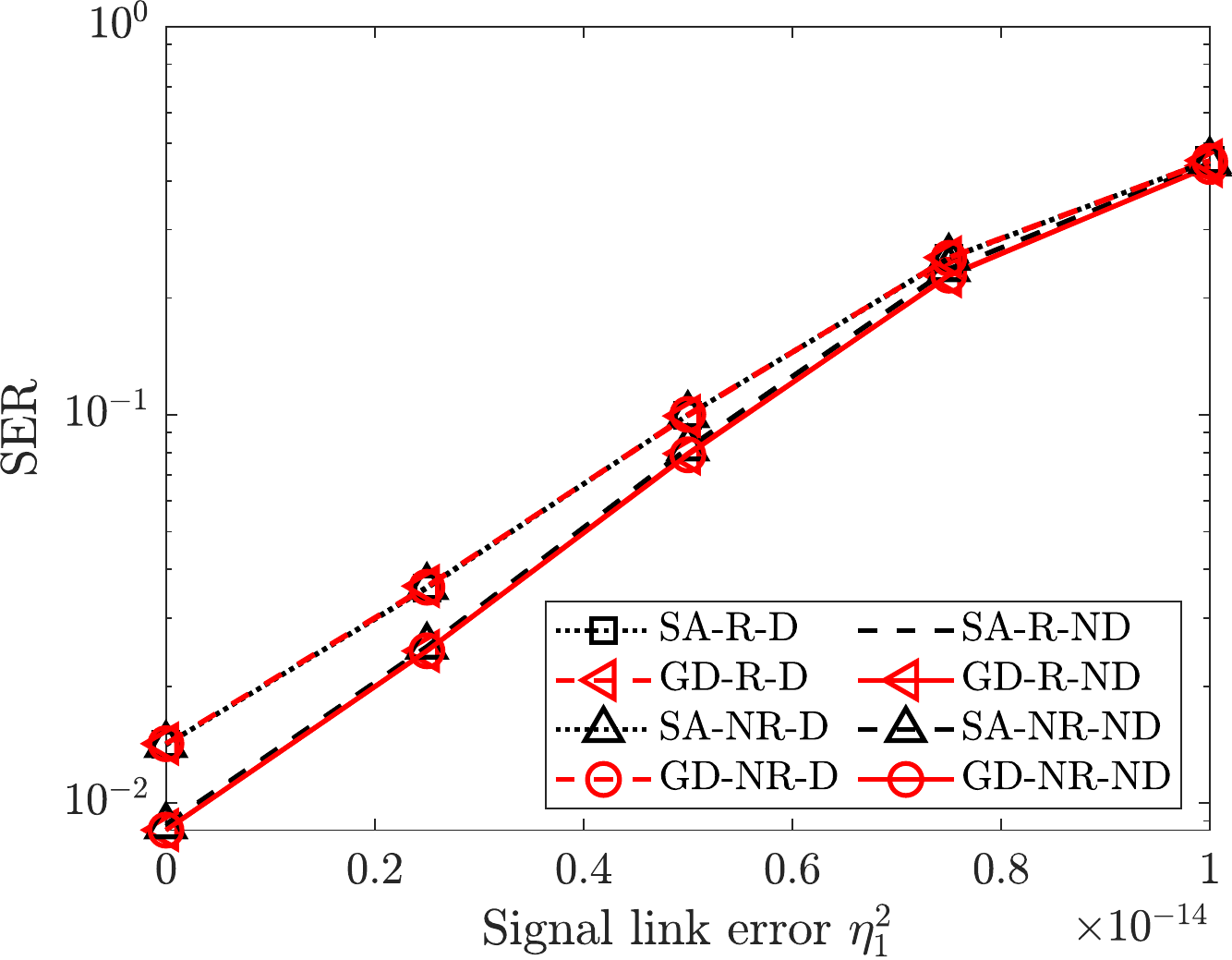}
         \caption{}
         \label{fig:e1}
     \end{subfigure}
     \begin{subfigure}[b]{0.32\textwidth} 
         \centering
         \includegraphics[width=\textwidth]{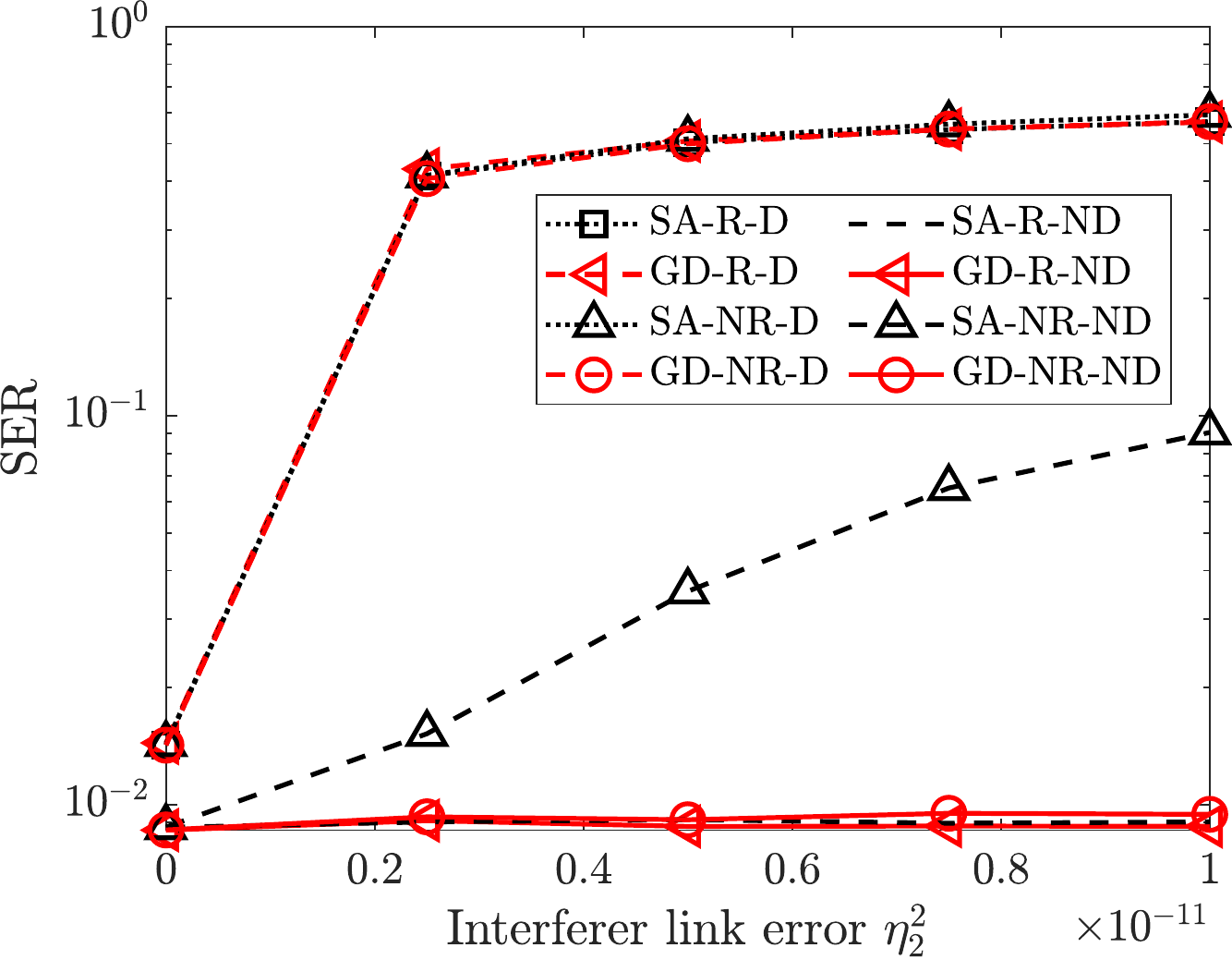}
         \caption{}
         \label{fig:e2}
     \end{subfigure}
                  \begin{subfigure}[b]{0.32\textwidth}
         \centering
         \includegraphics[width=\textwidth]{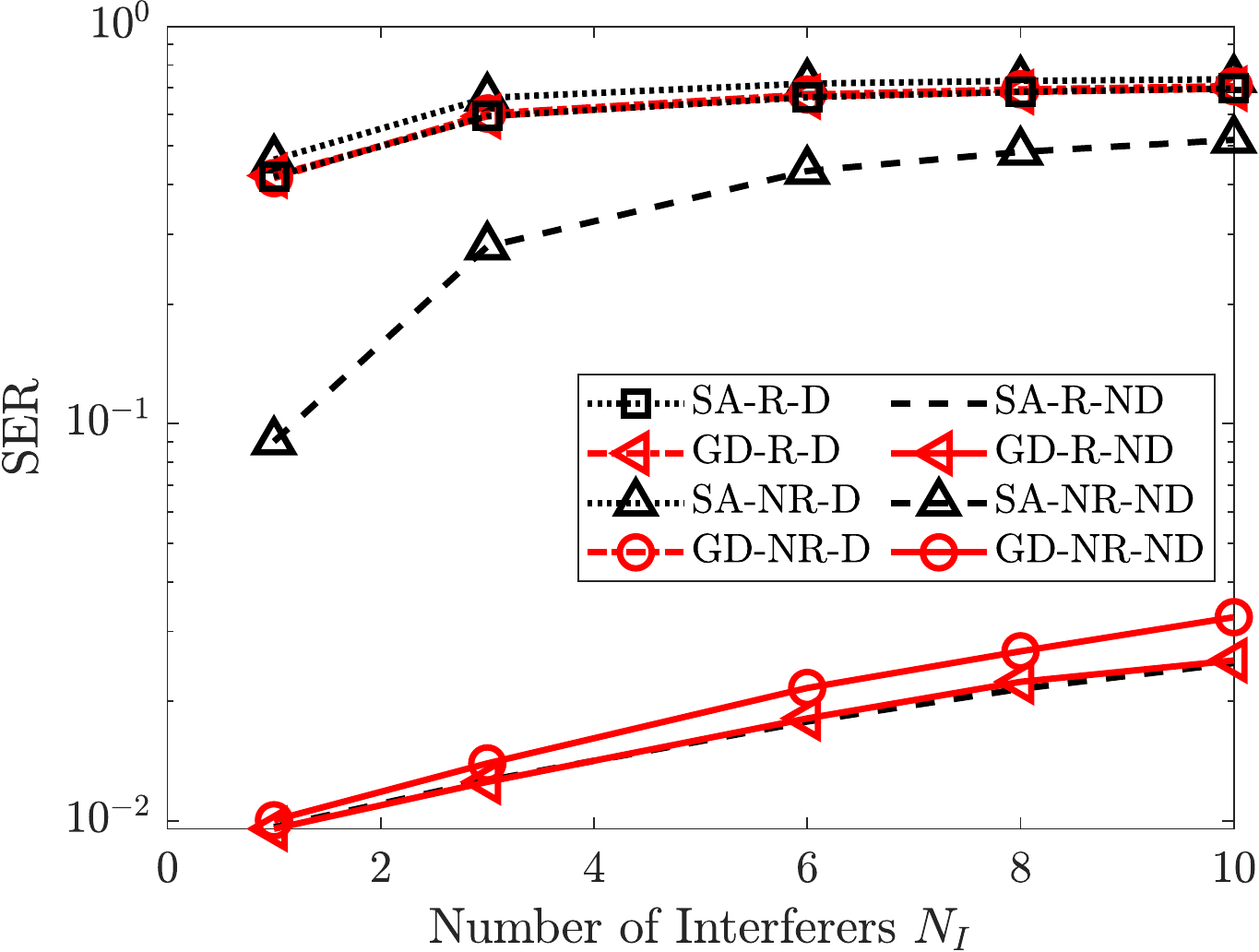}
         \caption{}
         \label{fig:Ni}
     \end{subfigure}
        \caption{Achievable throughput with the (a) normalized signal error amount, (b) normalized interferer error amount, and (c) SER with the number of interferers.}
        \label{fig:error}

\end{figure}

We plot SER versus CSI quality and number of interferers in Fig \ref{fig:error} where we consider CSI quality of the signal link and the interferer links by defining $\sigma_0^2=\eta_1^2$, and $\sigma_i^2=\eta_2^2$ for all $i=1,2,\hdots,N_I$, respectively. We also conduct the simulations under Assumption \ref{assumption:noise} and disregard BCD-SDR as it only acts as a computationally expensive baseline. In Fig. \ref{fig:e1}, we plot SER against $\eta_1^2$. The SER curves of both robust and non-robust algorithms overlap implying that the robust algorithms have limited benefits when signal link has error. In Fig. \ref{fig:e2}, we plot SER with interferer error $\eta_2^2$. When interferer direct links are not present, the SER of the non-robust BCD-SA algorithm is significantly worse than the other algorithms. The robust algorithms perform well in this scenario. In both these figures, SER performance worsens with increasing error amount while robust algorithms demonstrate limited SER improvement when interferer direct links are not blocked. Specifically, BCD-GD provides better SER performance over BCD-SA as the GD solution can help the RIS suppress interference unlike the SA solution where the RIS is only utilized to align the signal channels.

\subsubsection*{Effect of interferers with imperfect CSI on SER} 
For a more practical scenario, SER is plotted against the number of interferers under no signal error and high interferer channel error $(\eta_2^2=10^{-11})$ in Fig. \ref{fig:Ni}. This is a practical use-case when the interferers are non-cooperative and the Rx has limited information about the interferer channels. Further, the interferers are distributed randomly in a uniform manner on a circular ring of radius $2$ meter with ${\rm Rx}_0$ as the center. In this figure, SER performance of a robust algorithm is better than the non-robust counterpart in both the scenarios. We observe that the gap between them decreases with the number of interferers when direct links of the interferers are present, but exhibits increasing trend when direct links of the interferers are blocked. This emphasizes the usefulness of BCD-GD algorithm in the presence of non-cooperative interferers without direct links to the Rx.

\subsubsection*{Effect of transmission frequency on achievable throughput} Fig. \ref{fig:freq} shows the variation of achievable throughput with the frequency. As there is not much gap between assumptions in the `ND' scenario, we focus our discussion to the `D' scenario. In this case, the performance in Assumption \ref{assumption:scatter} is not affected much by the absorption coefficient peaks due to the NLOS nature of re-radiation unlike the performance in Assumption \ref{assumption:noise}. The performance loss in Assumption \ref{assumption:noise} results from the high molecular re-radiation noise in the absorption peaks.
\section{Conclusions}
In this work, we investigated the sensitivity of an RIS-aided THz system under two extreme manifestations of molecular re-radiation. In particular, we first developed a parametric THz channel model that accounts for both re-radiation assumptions through a simple parameter change. This channel model was then utilized to jointly optimize the RIS's phase-shift and receive beamformer with the objective of maximizing a lower bound on channel capacity that considers imperfect CSI. Specifically, we proposed an alternating BCD optimization framework that splits the original problem of two sets of variables (i.e., the RIS's phase shift and receive beamformer) into two sub-problems of a single set of variables each. These two sub-problems are then solved iteratively to converge to an efficient solution. In this framework, we approached the RIS sub-problem from three different directions: a) an SDR method that reformulates the sub-problem as a convex optimization problem while relaxing some constraints, and finds a near-optimal solution with high computational complexity, b) a fast SA method that maximizes the numerator of the SINR, and c) GD-based method that converges to a first-order stationary point of the original non-convex sub-problem. 

Our analytical results for a one-element RIS-aided system in the presence of a single interferer demonstrated that the SA solution is sub-optimal when the direct link of the interferer has comparable power to the reflected links. Several key system design insights were also obtained from our numerical results. For instance, our results revealed that the throughput of an optimized system is slightly higher in the scattering manifestation of re-radiation, and the exact difference depends on both the LOS probability of the direct links and frequency. On top of that, they showed that the peaks in the absorption coefficient for different frequencies have no additional impact on the throughput performance in the scattering manifestation of the re-radiation. Moreover, they demonstrated that the performance loss due to incorrect assumption of the re-radiation model in the optimization procedure is minimal under perfect CSI. Further, the throughput exhibits linear and logarithmic increasing trend against the number of RIS elements and Rx antennas, respectively. The results also demonstrated that the robust algorithms provide better SER when interferer direct links are blocked. Furthermore, they highlighted the efficacy of BCD-GD algorithm through its runtime and superior SER performance in the presence of non-cooperative interferers. To the best of our knowledge, this paper makes the first attempt to investigate the performance sensitivity of an optimized RIS-assisted THz system caused by different assumptions regarding molecular re-radiation. As the numerical and analytical results show that passive RIS has limited capability to combat powerful interference, using an active RIS is left as a promising direction for future work.
\appendix
\subsection{Proof of Lemma \ref{lem:NoiseRISdependence}} \label{sec:Lem1Proof}
From (\ref{eq:signalmodel0}), the signal power for the ${\rm Tx}_i$ along the direct path is $P_i\left(\frac{c}{4\pi f d_i}\right)^2$ and subsequently the molecular absorption noise variance due to the direct path would be $\sigma_{m_1,i}^2=\mathcal{I}_i\left(\frac{c}{4\pi f d_i}\right)^2P_i[1-\tau(f,d_i)]$. The indicator function ensures that the noise exists only when the direct link is present.

If the distances between ${\rm Tx}_i$ to RIS, and RIS to ${\rm Rx}_0$ are $d_{\gamma_i}$ and $d_\alpha$, respectively, we inspect the signal $x$ of power $P_i$ through the $m$-th element of RIS with reflection coefficient $(\alpha_me^{j\theta_m})$ without including path-loss terms for simplicity.

The incident signal on the RIS is $x\sqrt{\tau(f,d_{\gamma_i})}+n_1$ where $n_1\sim\mathcal{CN}(0,P_i(1-\tau(f,d_{\gamma_i})))$ is the additive molecular absorption noise. Ultimately, the reflected signal from RIS is
\begin{align*}
  y=&(x\sqrt{\tau(f,d_{\gamma_i})}+n_1)\alpha_me^{j\theta_m}\sqrt{\tau(f,d_{\alpha})}+n_2.
\end{align*}
As the reflected power from the RIS element is $|\alpha_m|^2P_i$, $n_2\sim\mathcal{CN}(0,|\alpha_m|^2P_i(1-\tau(f,d_{\alpha})))$ is the additive noise for the RIS to ${\rm Rx}_0$ path. So, the noise variance due to both the paths is
\begin{align*}
    &\mathrm{E}[|n_1\alpha_me^{j\theta_m}\sqrt{\tau(f,d_{\alpha})}+n_2|^2]\\
   &=|\alpha_m|^2\tau(f,d_{\alpha})P_i(1-\tau(f,d_{\gamma_i}))+|\alpha_m|^2P_i(1-\tau(f,d_{\alpha})) \\
   &=P_i|\alpha_m|^2[1-\tau(f,d_\alpha)\tau(f,d_{\gamma_i})].
\end{align*}

Extending this result to an $N$-element RIS, if the RIS-${\rm Rx}_0$ and ${\rm Tx}_i$-RIS channels are $\mathbf{a}_{h_1}$, and $\mathbf{a}_{h_2}$ with their entries as array factors $a_{h_{1,m}}$ and $a_{h_{2,m}}$ with ULA assumption for RIS, the received signal for SISO is
\begin{align*}
   y
   \!=\!\!x\sqrt{\tau(f,d_{\gamma_i})\tau(f,d_{\alpha})} \!\!\sum\limits_{m=1}^N \!\!\left(\!\alpha_me^{j(\theta_m+a_{h_{1,m}}+a_{h_{2,m}})}\!\right) \!\!+\!\!\! \sum\limits_{m=1}^N\!\! n_m,
\end{align*}
where $\sum_{m=1}^Nn_m\sim\mathcal{CN}(0,P_i[1-\tau(f,d_\alpha)\tau(f,d_{\gamma_i})]\sum_{m=1}^N|\alpha_m|^2)$. By including path-loss terms, and writing $\sum_{m=1}^N|\alpha_m|^2$ in matrix form, the molecular noise variance for the reflected signal through RIS can be written as $\sigma_{m_2,i}^2\mathbf{\boldsymbol{\theta}}^H\mathbf{\boldsymbol{\theta}}$ where $\sigma_{m_2,i}^2=\left(\frac{c^2}{16(\pi f)^2 }\frac{1}{d_\alpha d_{\gamma_i}}\right)^2P_i[1-\tau(f,d_\alpha)\tau(f,d_{\gamma_i})]$. The molecular absorption noise variance is then $\zeta\sigma_{m,i}^2$, where $\sigma_{m,i}^2=\sigma_{m_1,i}^2+N\sigma_{m_2,i}^2$ as this noise will only exist for Assumption \ref{assumption:noise} or $\zeta=1$.

\subsection{Proof of Theorem \ref{lem:SPsolution}} \label{sec:TheoProof}
We start with the generic SINR expression:
\begin{align}
    \gamma_1=\frac{L'+M'\cos(s+x)}{N'+P'\cos(t+x)}. \label{twoSINRproof}
\end{align}
Differentiating \eqref{twoSINRproof} with respect to $x$ and equating it to zero results in the following equations:
\begin{align}
    &\frac{L'+M'\cos(s+x)}{N'+P'\cos(t+x)}=\frac{M'\sin(s+x)}{P'\sin(t+x)} \label{eq:SP1}, \\
    &\frac{\sin(t+x)}{M'N'}-\frac{\sin(s+x)}{L'P'}=\frac{\sin(s-t)}{L'N'} \label{eq:SP2}.
\end{align}
\eqref{eq:SP2} can be derived from \eqref{eq:SP1}. However, \eqref{eq:SP1} provides an alternate form of $\gamma_1$ at a stationary point. Using \eqref{eq:SP1}, the SINR can be expressed as a function of $\sin(t+x)$:
\begin{align}
    \gamma_1=\frac{L'}{N'}-\frac{M'\sin(s-t)}{N'\sin(t+x)}. \label{eq:realtwo}
\end{align}
Note that, \eqref{eq:SP2} can be expressed as a quadratic equation of $\sin(t+x)$. The solutions for $\sin(t+x)$ plugged in \eqref{eq:realtwo} proves the theorem.

\hfill 
\bibliographystyle{IEEEtran}
\bibliography{hokie_v1}

\end{document}

%% file: notation.tex
\def\nb0{{\mathbf{0}}}
\def\nb1{{\mathbf{1}}}

\newtheorem{lemma}{Lemma}

\newtheorem{theorem}{Theorem}

\newtheorem{cor}{Corollary}

\newtheorem{remark}{Remark}
\newtheorem{assumption}{Assumption}
	
